\documentclass[aps,prx,twocolumn,superscriptaddress,nofootinbib]{revtex4-2}
\usepackage[utf8]{inputenc}
\usepackage[T1]{fontenc}
\usepackage{braket,color}
\usepackage{graphicx,enumerate,verbatim,bbold}
\usepackage{amsmath,amssymb,amsthm,mathrsfs}
\usepackage{dsfont}
\usepackage{lmodern}
\usepackage{quantikz}
\usepackage{braket}
\usepackage{subfigure}
\usepackage{subcaption}
\usepackage{multirow}
\usepackage{physics}
\usepackage{bm}
\usepackage[normalem]{ulem}
\usepackage[caption=false]{subfig}
\usepackage{dcolumn}
\newcommand{\bqa}{\begin{eqnarray}}
\newcommand{\eqa}{\end{eqnarray}}

\newcommand{\be}{\begin{equation}}
\newcommand{\ee}{\end{equation}}

\newcommand{\beginsupplement}{%
        \setcounter{table}{0}
        \renewcommand{\thetable}{S\arabic{table}}%
        \setcounter{figure}{0}
        \renewcommand{\thefigure}{S\arabic{figure}}%
        \setcounter{equation}{0}
        \renewcommand{\theequation}{S\arabic{equation}}%
     }
\begin{document}
\title{Multi-ion entangling gates
mediated by spectrally unresolved modes
}

\author{Modesto Orozco-Ruiz}
\affiliation{Physics Department, Blackett Laboratory, Imperial College London, Prince Consort Road, SW7 2AZ, United Kingdom}
\author{Florian Mintert}
\affiliation{Physics Department, Blackett Laboratory, Imperial College London, Prince Consort Road, SW7 2AZ, United Kingdom}

\begin{abstract}
Entangling interactions between distant qubits can be mediated via an additional degree of freedom. In conventional trapped-ion schemes, realizing a well-defined, coherent gate typically requires spectrally addressing a specific bus mode. As the ion number increases, the coupling to each individual motional mode becomes weaker, so gates on large ion strings mediated by a single mode are necessarily slow. Moreover, addressing a large number of modes demands complex driving schemes, and the fundamentally perturbative character of these approaches imposes constraints on achievable gate speed and fidelity.

Here, we introduce a scheme for entangling trapped-ion qubits using a time-dependent magnetic-field gradient, in which all axial motional modes participate in mediating the interaction and the gate construction is nonperturbative. The framework can be used to implement both multi-qubit gates and two-qubit gates between arbitrary pairs in a linear ion string. Through several explicit examples, we highlight the advantages over existing magnetic-gradient schemes and show how gates on multiple ion pairs can be carried out simultaneously.

\end{abstract} 
\maketitle

\section{Introduction}
Trapped ions provide one of the most prominent platforms for quantum information processing, in which entangling interactions are mediated via a degree of freedom that is not used to encode the qubits themselves. Owing to their mutual Coulomb repulsion, any intrinsic interaction between the ions' electronic degrees of freedom is practically negligible, and entangling gates must instead be implemented via the strongly interacting motional degrees of freedom~\cite{bruzewicz2019trapped}.

State-of-the-art realizations of entangling interactions are based on spectral selection of a specific motional mode~\cite{sorensen1999quantum}. The coupling of each ion to a given mode decreases with increasing number of trapped ions~\cite{haffner2008quantum, leung2018entangling}, so implementing a gate at a fixed speed requires stronger driving for longer ion chains. Since spectral addressing is fundamentally a perturbative, weak-driving principle, there is an inherent tension between fast gates, which demand strong driving, and high-fidelity gates, which demand weak driving, and this tension becomes more pronounced as the number of trapped ions increases. Consequently, most high-fidelity gates demonstrated to date have involved only two ions confined in a single potential well~\cite{bruzewicz2019trapped}.

Most architectures for actual computing devices, however, have trapping zone that hold larger numbers of ions~\cite{kielpinski2002architecture}.
Experiments with ions strings of up to a few dozens of ions are not uncommon anymore~\cite{ransford2025helios, pagano2018cryogenic}, but most of the gate schemes in use have been designed with short ion strings in mind.

Here, we describe a control scheme for realizing entangling interactions that does not require spectral addressing and that, in general, exploits all of the ions' motional modes. This work builds on microwave-driven gates that use magnetic-field gradients to induce qubit-state-dependent forces~\cite{mintert2001ion, ospelkaus2011microwave}.
When implemented with surface-electrode traps, such approaches allow for compact integration and potentially improved scalability~\cite{chiaverini2005surface, lekitsch2017blueprint}. 
Unlike their optical counterparts, microwave-driven gates do not necessarily require the Lamb-Dicke approximation and can tolerate higher motional excitation~\cite{mintert2001ion}.
However, in practical implementations, the magnetic field gradients that can be realized are limited, and the resulting qubit-motion coupling is typically weaker than what can be achieved with optical fields.
As a result, gate times are generally longer~\cite{bruzewicz2019trapped, harty2016high, weidt2016trapped, schafer2018fast}.

The strength of the qubit-motion coupling is proportional to the magnetic field gradient. 
Since the effective interaction between two qubits is a second order process, its strength is quadratic in the magnetic field gradient. A significant enhancement of the qubit-qubit interaction can be achieved by modulating the magnetic field gradient at a frequency close to one of the motional mode frequencies~\cite{ospelkaus2008trapped}.
In this regime, the effective interaction is amplified by a factor of $1/\delta$, where $\delta$ is the detuning between the modulation frequency and the targeted motional mode.

This resonant enhancement approach, however, is not without its limitations. Since the driving fields are tuned close to the frequency of a specific motional mode, they must be sufficiently detuned from other modes to avoid crosstalk. Under the rotating-wave approximation used to derive the effective interaction, rapidly oscillating terms are neglected. While this approximation is valid for sufficiently small detunings, it restricts the range of permissible driving parameters and introduces a trade-off between gate speed and fidelity.
As the coupling of a qubit to one motional mode decreases roughly as $1/\sqrt{N}$, fast gates in long chains often require participation of multiple motional modes, making mode-selective approaches less effective.

Our present work aims at overcoming these limitations. Rather than avoiding the influence of off-resonant modes, our approach is designed to \emph{harness} their collective dynamics.
By doing so, we engineer qubit–qubit interactions mediated by \textit{all} motional modes, enabling robust and scalable entangling operations that remain effective even as the number of ions increases.

This collective-mode strategy overcomes two central limitations of conventional gate schemes: the need to spectrally resolve individual motional modes, and the diminishing coupling strength per mode as the system grows. Moreover, since our derivation makes no use of perturbative approximations such as rotating wave approximations, the resulting dynamics remain valid for strong gradients.
In this way, the gate speed is fundamentally limited only by the maximum magnetic field gradient achievable in the experiment.

\section{Framework}

The system we consider consists of $N$ ions confined in a common potential well and coupled to $N$ collective motional modes via a time-dependent magnetic field gradient. The Hamiltonian describing this system is given by
\begin{align}
H(t) =\sum_{j=1}^N\frac{\omega_j}{2}Z_j+\sum_{l=1}^N \nu_la_l^\dagger a_l+f(t)\sum_{j,l} \nu_l \eta_{jl}Z_j(a_l^\dagger+a_l)\ ,
\label{eq:H}
\end{align}
where $\omega_j$ is the qubit frequency for ion $j$, and $Z_j$ denotes the corresponding Pauli-$Z$ operator. The operators $a_l^\dagger$ and $a_l$ create and annihilate excitations in the $l$-th normal mode of motion with frequency $\nu_l$. The dimensionless coupling strength $\eta_{jl}$ between ion $j$ and motion in normal mode $l$ is given by
\begin{align}
    \eta_{jl} = g_F m_F \mu_B   \frac{z_0^{(l)}}{ \hbar \nu_l} 
    \chi_{jl} \frac{\partial B}{\partial z} \ ,
\label{eq:eta}
\end{align}
where the scalars $\chi_{jl}$ describe the displacement of ion $j$ in terms of the normal modes $l$, $\mu_B$ is the Bohr magneton, $g_F$ is the hyperfine Landé $g$-factor, $m_F$ is the magnetic quantum number, and $z_0^{(l)} = \sqrt{\hbar / (2m \nu_l)}$ is the zero-point extent of mode $l$ for an ion of mass $m$. The magnetic field gradient $\frac{\partial B}{\partial z}$ is assumed to be uniform across the ion chain.

Time-dependent modulations of the magnetic field gradient are described by the envelope function $f(t)$ (such that $| f(t)|\le 1$ at any time $t$) 
in Eq.~\eqref{eq:H}, which controls the temporal profile of the interaction strength. For example, $f(t)=1$ corresponds to a static gradient, while a monochromatic drive at frequency $\omega$ would correspond to $f(t) = \cos(\omega t)$. The present approach exploits more general time-dependent profiles of $f(t)$ in order to \textit{intentionally} excite and manipulate the full collective motion of the ion chain.

\subsection{Effective interactions}

Similar to the case of a static field gradient, the Hamiltonian can be diagonalized with a polaron transformation,
but a time-dependent gradient requires a time-dependent transformation.
With the specific choice
\be
U_P(t) = \exp( i \sum_{j,l} \left( g_l^*(t) a_l + g_l(t) a_l^{\dagger} \right)\eta_{jl} Z_j)\ ,
\label{eq:polaron}
\ee
where the time-dependent functions $g_l(t)$ satisfy the equations of motions
\be
\dot g_l(t) +i \nu_l g_l(t)  = \nu_l f(t)
\label{eq:eqs_motion}
\ee
of a driven harmonic oscillator,
the transformed Hamiltonian $\tilde H(t) = U_P(t) H(t) U_P^{\dagger}(t) + i\dot U_P(t) U_P^{\dagger}(t)$ reads
\be
\tilde H(t)= \sum_j\frac{\omega_j}{2}Z_j+\sum_l\nu_la_l^\dagger a_l+\sum_{ij}\alpha_{ij}(t)Z_i Z_j
\label{eq:Hpolaron}
\ee
with coupling elements $\alpha_{ij}(t) =\sum_{l} \eta_{il} \eta_{jl} \Phi_l(t)$ given in terms of the real functions
\begin{align}
    \Phi_l(t) &= \nu_l f(t) \mathfrak{Im}(g_l(t))
    \ ,
    \label{eq:phases}
\end{align}
with the $\mathfrak{Im}(g_l(t))$ denoting the imaginary component of $g_l(t)$.

Although the Hamiltonian $\tilde H(t)$ is generally time-dependent, the commutator $[\tilde H(t_1),\tilde H(t_2)]=0$ holds for all $t_1$ and $t_2$. This property ensures that the time-evolution operator $\tilde U(t)$ induced by $\tilde H(t)$ is of the simple form 
$\tilde U(t)=\exp(-i\int_{0}^td\tau\tilde H(\tau))$,
eliminating the need for time-ordering. The dynamics generate geometric phases $\int_{0}^Tdt\Phi_l(t)$, that depend on the phase-space trajectory of the oscillator mode $l$.

Given the ability to implement a suitable driving function $f(t)$, the dynamics induced by the Hamiltonian $\tilde H(t)$ in Eq.~\eqref{eq:Hpolaron} generate mode-dependent phase shifts
\begin{align}
    D_l = \int_{0}^{T} dt \, \Phi_l(t) \ .
\label{eq:phase_shifts}
\end{align}
Collecting these into the diagonal matrix $D = \mathrm{diag}(D_1,\dots,D_N)$, the propagator $\tilde U(T)$ at time $T$ takes the form
\begin{equation}
    \tilde U(T)
    = \exp[-i\left(\tilde H_0 T + \sum_{j\neq k} \Lambda_{jk} Z_j Z_k\right)] \ ,
\end{equation}
with a non-interacting contribution $\tilde H_0$ and an interaction matrix $\Lambda$ whose elements are
\begin{equation}
    \Lambda_{jk} = \sum_{l} \eta_{jl} \eta_{kl} D_l \, ,
\label{eq:alpha_star}
\end{equation}
\textit{i.e.}, $\Lambda = \eta D \eta^\top$.

Since $Z_j^2 = \mathbb 1$ for all $j$, any diagonal term of the form $\Lambda_{jj} Z_j^2$ contributes only a global phase and can be absorbed into $\tilde H_0$ without affecting the entangling action of the gate. Up to such global phases, the effective entangling interaction is therefore fully specified by the off-diagonal entries of $\Lambda$, which are determined by the $N$ mode-dependent phases $D_l$ set by the temporal modulation $f(t)$ of the magnetic-field gradient. For a system of $N$ qubits, the global modulation $f(t)$ thus provides $N$ independent tunable parameters with which to synthesize an interaction geometry.

By contrast, the most general geometry of pairwise $ZZ$ interactions corresponds to an arbitrary real symmetric matrix with vanishing diagonal, specified by $N(N-1)/2$ independent parameters. The restriction to only $N$ tunable parameters in our setting arises from the constraint to a global field gradient. In principle, this limitation could be overcome by employing local, individually controllable gradients at each ion, but realizing strong, spatially varying magnetic fields on the scale of the ion--ion separation in a strongly confining trap is extremely challenging experimentally. In the following, we therefore maintain the constraint of a global magnetic-field gradient and instead invoke additional spin-echo techniques to realize the most general interaction geometry of pairwise $ZZ$ couplings.

\subsection{Spin echo}\label{sec:spin_echo}

Interaction matrices $\Lambda$ that are not of the form of Eq.~\eqref{eq:alpha_star} can be realized in combination with spin echo techniques~\cite{hahn1950spin,carr1954effects}\,---\,a technique that is well established in the context of static magnetic‐field gradients~\cite{wang2009individual, harty2016high}.

This is based on the fact that applying $\pi$-pulses $\exp(-i\frac{\pi}{2}X)$ to a selected subset of ions before evolution under $\tilde H(t)$ in Eq.~\eqref{eq:Hpolaron}, and the corresponding inverse pulses $\exp(i\frac{\pi}{2}X)$ afterwards, produces dynamics equivalent to evolution under a Hamiltonian of the same form but with the vector $\vec{\omega}$ of resonance frequencies replaced by $O\vec{\omega}$,
and the interaction matrix $\alpha$ replaced by $O\alpha O^{\top}$,
where the diagonal matrix $O$ has entries $-1$ for ions addressed by the $\pi$-pulses and $+1$ for any other ion.

A propagator of the form
\be
\exp\left(i\sum_{jk}\Lambda_{jk}Z_jZ_k\right)
\ee
can thus be turned into the propagator
\be
\exp\left(i\sum_{jk}[O\Lambda O^{\top}]_{jk}Z_jZ_k\right)
\ee
with a spin echo.

Since all propagators of this form commute with each other, a sequence of $N_I$ periods of controlled dynamics interleaved with $\pi$-pulses thus yields the effective interaction matrix
\begin{align}
    \Lambda = \sum_{p=1}^{N_I} O^{(p)} \Lambda^{(p)} (O^{(p)})^{\top} \ ,
\label{eq:alphaT_sequence}
\end{align}
where $\Lambda^{(p)}$ is the interaction matrix for period $p$, and $O^{(p)}$ is the matrix for the corresponding $\pi$-pulses.
This lifts the restriction to the interaction matrices in Eq.~\eqref{eq:alpha_star}, and the use of sufficiently many periods gives access to the most general interaction matrix $\boldsymbol{\varLambda}$ that one might aim at realizing.

The construction of elementary interaction matrices $\Lambda^{(p)}$ and corresponding $\pi$-pulses can be performed for any targeted gate.
As shown in the following, the realization of an interaction of exactly one pair of ions can be achieved with $N_I=2$ periods of controlled dynamics.
This construction directly yields any desired interaction geometry as sequence of pairwise interactions, but bespoke constructions will likely yield constructions with fewer periods of controlled dynamics.

An interaction matrix $\boldsymbol{\varLambda}$ with a finite interaction for exactly one pair of ions can be obtained with two elementary interaction matrices $\Lambda^{(p)}$ satisfying $\Lambda^{(2)}=-\Lambda^{(1)}$.
There are no $\pi$-pulses for the first period of controlled dynamics, but the second period is dressed by a $\pi$-pulse on ion $k$, such that the matrix $O^{(2)}$ has the elements $O^{(2)}_{kk}=-1$ and $O^{(2)}_{jj}=1$ for $j\neq k$. With the transformation
\be
\big[O^{(2)}\Lambda^{(2)}(O^{(2)})^{\top}\big]_{ij}
=\Lambda^{(2)}_{ij}
\ee
for both $i$ and $j$ different than $k$,
and for both $i$ and $j$ equal to $k$,
and
\be
\big[O^{(2)}\Lambda^{(2)}(O^{(2)})^{\top}\big]_{ik}
=
\big[O^{(2)}\Lambda^{(2)}(O^{(2)})^{\top}\big]_{ki}
=-\Lambda^{(2)}_{ik}
\ee
for $i\neq k$,
the total interaction matrix reads
\begin{align}
\nonumber \Lambda &= 
\Lambda^{(1)}+O^{(2)}\Lambda^{(2)}\big(O^{(2)}\big)^{\top} \\
&= 2\sum_{j, j\neq k} \Lambda^{(1)}_{kj}\,\big(\dyad{k}{j} + \dyad{j}{k}\big)\ .
\label{eq:alphaT}
\end{align}
This corresponds to an interaction geometry in which ion $k$ interacts with any other ion, but the ions different than $k$ do not interact with each other.

The remaining freedom lies in the choice of $\Lambda^{(1)}$, which is still parametrized by the $N$ mode-dependent phases $D_l$ (following Eq.~\eqref{eq:alpha_star}), determined by the modulation of the field gradient via Eq.~\eqref{eq:phase_shifts}. These $N$ free parameters are sufficient to realize a finite value for the interaction constant between ion $k$ and some other selected ion while realizing a vanishing interaction between ion $k$ and any other ion. Two periods of controlled dynamics combined with a simple spin echo thus suffice to realize a pure pairwise $ZZ$ coupling on a chosen ion pair.

\subsection{Boundary conditions}\label{sec:boundary_conds}

The polaron transformation Eq.~\eqref{eq:polaron} describes the concept of dressing the bare qubit degrees of freedom with the motion.
While the Hamiltonian $H(t)$ in Eq.~\eqref{eq:H} captures the interaction of the bare qubits with the motional degrees of freedom, the Hamiltonian $\tilde H(t)$ in Eq.~\eqref{eq:Hpolaron} captures the interactions among the dressed qubits.

For the dressed-qubit picture to remain consistent throughout a sequence of several gates, one must impose boundary conditions on $g_l(t)$ at the start ($t=0$) and end ($t=T$) of each gate.
The only requirement is that a gate’s final condition coincides with the next gate’s initial condition\,---\,but how one chooses those conditions depends on the gradient scheme in use.

To enable a direct comparison between the schemes developed here and earlier approaches employing static gradients~\cite{mintert2001ion} or oscillating gradients~\cite{ospelkaus2008trapped, ospelkaus2011microwave}, we next discuss boundary conditions that are consistent with these two established methods.

\subsubsection{Consistency with static gradient}\label{sec:static_bc}

In the case of a static gradient, qubits are defined by dressed eigenstates in the presence of an always‐on field gradient. Initialization and readout therefore occur with a nonzero gradient at both $t=0$ and $t=T$. Consistent with this picture, the boundary conditions can be chosen as
\be
ig_l(0)-f(0)=ig_l(T)-f(T)=0\ ,
\label{eq:condit_static}
\ee
which, from Eq.~\eqref{eq:eqs_motion}, implies $\dot{g}_l(0) = \dot{g}_l(T)$.
These conditions guarantee that the dressed‐qubit basis at $t=0$ and $t=T$ matches the static‐gradient eigenbasis, and that the ions motion is not altered as a result of the gate operation. 

A key distinction, however, is that in the present scheme $f(t)$ is time‐dependent, so $f(0)$ and $f(T)$ need not be equal. As a result, the initial polaron transformation $U_P(0)$ may differ from the final transformation $U_P(T)$, allowing the collective motional states to be displaced by different amounts at the start and end of the gate while remaining stationary at both times.

\begin{figure*}[t]
    \centering \includegraphics[width=0.99\textwidth]{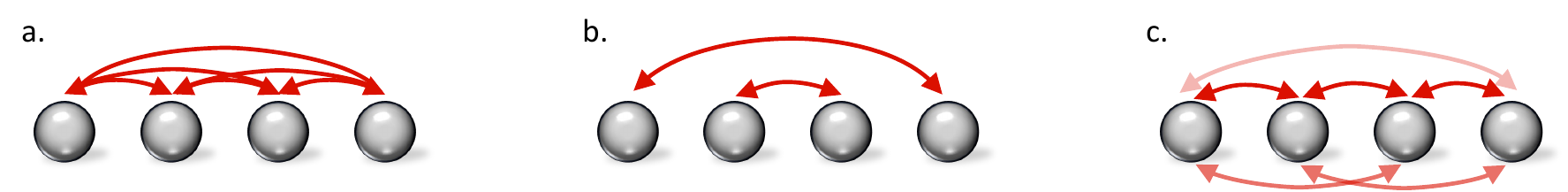}
    \caption{Illustration of various interaction geometries, where arrows indicate interacting elements, and their color represent interaction strength (light for weak, and dark for strong). In (a), uniform coupling is shown across all ions, with equal interaction strength between each pair. In (b), a rainbow-style entanglement geometry is depicted, where symmetric pairs of ions are coupled with equal strength. In (c), distance-dependent coupling is represented, where the ion at site $j$ interacts with ions at sites $k>j$, with interaction strength decreasing as a function of the distance between them. This interaction pattern forms the basis of the controlled-phase gates used in implementing the quantum Fourier transform.}
    \label{fig:geometries}
\end{figure*}

\subsubsection{Consistency with oscillating gradients}\label{sec:oscill_bc}

In the case of oscillating gradients, gates are implemented using finite gradients that are switched on and off instantaneously at the start and end of the gate operation, so that outside the gate window ($t<0$ or $t>T$) the gradient vanishes and qubits reduce to pure spin states. 

Thus, to decouple the qubit‐qubit entangling interaction from the initial phonon state, one enforces
\begin{align}
g_l(T) = e^{-i T \nu_l} g_l(0) \ ,
\label{eq:boundary_alpha}
\end{align}
which ensures that the net motional displacement vanishes at the end of the gate. This cleanly isolates the spin dynamics and yields a spin-spin entangling operation free from phonon-state dependence.

\subsection{Gate synthesis}\label{sec:gate_synthesis}

Building on Sec.~\ref{sec:spin_echo}, the synthesis of a general $N$-qubit entangling gate with prescribed interaction
\begin{align}
    \sum_{j,k}\boldsymbol{\varLambda}_{jk} Z_j Z_k \, ,
\end{align}
where $\boldsymbol{\varLambda}$ is the target coupling matrix, typically relies on a sequence of global driving segments $f_m(t)$, interleaved
with layers of $\pi$ pulses. For each segment $m$ we denote by
$U_{\pi}^{(m)}$ the product of $\pi$-pulses applied immediately before and
after the corresponding gradient–modulation $f_m(t)$, and by
$\tilde U[f_m]$ the propagator generated during that interval under $\tilde H(t)$. The net gate can then be written as 
\vspace{0.2em}
\begin{align}
\tilde U(T)
= \prod_m U_\pi^{(m)} \, \tilde U[f_m]\, U_\pi^{(m)} \, ,
\end{align}
with the product understood to be time-ordered.

Gate synthesis in this setting is naturally formulated as a mixed
discrete–continuous optimization problem. The continuous degrees of freedom are
the waveforms $f_m(t)$ and segment durations $T_m$, while in each interval $m$
there is, for every ion, a discrete choice of whether a $\pi$-pulse is applied
or not.

In practice, one can use an iterative procedure that jointly optimizes the
$\pi$-pulse patterns, the drives, and the gate times; the concrete strategy is
described in Sec.~II of the Supplemental Material~\cite{supp}. The role of the $\pi$-pulses is to
extend the class of effective couplings that can be realized, whereas the
drives $f_m(t)$ are shaped so as to approach the target pairwise phase shifts in Eq.~\eqref{eq:phase_shifts} while simultaneously satisfying the boundary conditions discussed in
Sec.~\ref{sec:boundary_conds}. Enforcing these via either
Eq.~\eqref{eq:condit_static} or Eq.~\eqref{eq:boundary_alpha} amounts to
imposing two real constraints per mode (for the real and imaginary parts),
\textit{i.e.} a total of $2N$ real conditions. For the parametrization in
Eq.~\eqref{eq:parametrization}, these appear explicitly as
Eqs.~(S2a) and~(S2b) in the Supplemental Material~\cite{supp}.

It is therefore natural to employ an ansatz for $f_m(t)$ with a number of free
parameters that scales at least linearly with the number of qubits (and modes)
$N$: one needs at least $2N$ tunable real coefficients to satisfy the boundary conditions, plus additional flexibility to shape the interaction. In the explicit constructions of Sec.~\ref{sec:gates} we use $2N+1$ basis functions
for $f_m(t)$, each specified by a tuple $(A_j,\omega_j,\vartheta_j)$, so that each segment is described by $3(2N+1)$ real parameters. This choice is
sufficient to obtain high-fidelity gates for small registers. For larger
systems (see Sec.~\ref{sec:largesystem}), further reductions in infidelity can
be achieved by enlarging the parametrization, but in practice we find that
keeping the total number of free parameters proportional to $N$ still suffices, and the optimization remains scalable.

\section{Quantum gates}
\label{sec:gates}

Finding an explicit driving protocol, including temporal modulation of the magnetic field and application of $\pi$-pulses according to Eq.~\eqref{eq:alphaT_sequence} does not require any simulation of the actual system dynamics.
As such, this is \emph{not} limited by the exponential scaling of Hilbert space with the number of qubits and motional modes.

An explicit analysis of the motional dynamics during a gate operation and comparison to existing gate schemes, however, does require such simulations.
This section is thus focused on a system with four ions that admit a detailed analysis of the full dynamics; gate design for larger systems can be found further down in Sec.~\ref{sec:largesystem}.

To demonstrate the methodology, we consider a time-dependent magnetic field gradient parametrized as
\begin{align}
f(t) &= \sum_j A_j \cos(\omega_j t + \vartheta_j) \ ,
\label{eq:parametrization}
\end{align}
with real amplitudes $A_j$, frequencies $\omega_j$ and phases $\vartheta_j$. As discussed in Sec.~\ref{sec:gate_synthesis}, any ansatz must provide enough degrees of freedom to satisfy the phase-space boundary conditions and, at the same time, to optimize the temporal profile towards the desired interaction geometry. The expansion in Eq.~\eqref{eq:parametrization} meets these requirements and is used throughout our numerical examples, since it leads to compact analytical expressions that facilitate the optimization (see Sec.~I of the Supplemental Material~\cite{supp} for details).

In the case of four ions, a decomposition involving nine frequency components, \textit{i.e.}, nine sets of parameters $(A_j, \omega_j, \vartheta_j)$, is typically sufficient to meet the boundary conditions (Eq.~\eqref{eq:condit_static} or Eq.~\eqref{eq:boundary_alpha}) and to shape the evolution toward the desired outcome. More efficient parameterizations may exist, but this choice already provides ample flexibility for tailoring the interaction. With this decomposition, one can engineer a wide range of effective couplings of the form \(\sum_{i,j} \Lambda_{ij} Z_i Z_j\), enabling control over the emergent qubit–qubit interactions.

\begin{figure*}[t]
    \centering \includegraphics[width=0.9\textwidth]{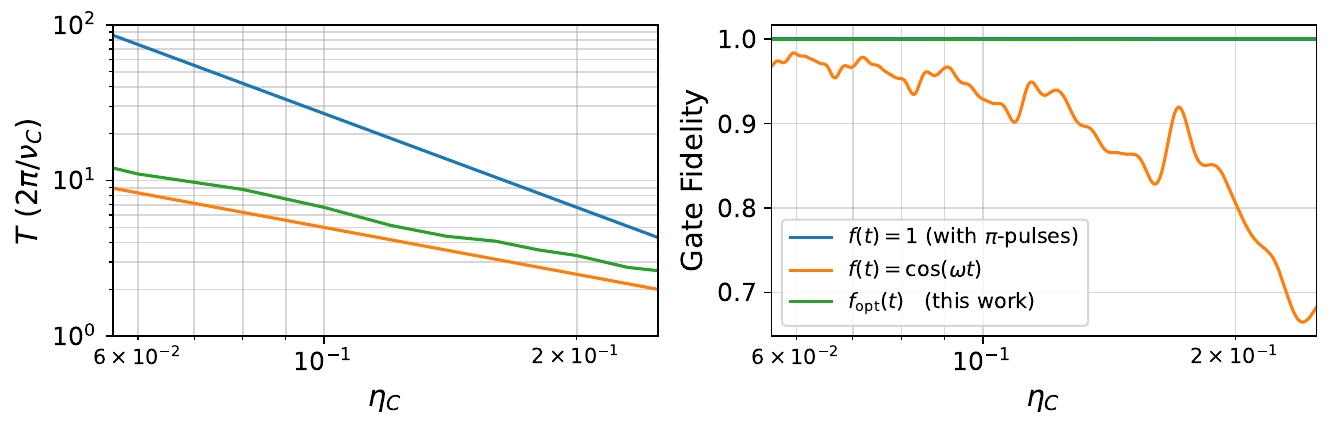}
    \caption{(a) Required gate duration, expressed in units of the COM mode period, as a function of the Lamb-Dicke parameter for the COM mode (values for other modes follow accordingly), for a static magnetic field gradient (blue) a monochromatic oscillating gradient (orange), and the optimized drive $f_{\text{opt}}(t)$ from this work (green). (b) Corresponding gate fidelities for the three protocols as a function of the Lamb-Dicke parameter, highlighting the trade-off between speed and fidelity in each case. All results correspond to a four-ion chain and the generation of an effective long-range Ising interaction as in Eq.~\eqref{eq:Hising}, with target coupling strength $J=\pi/4$. Fidelity is computed as described in Sec.~IV of the Supplemental Material~\cite{supp}.}
    \label{fig:comparison_static_monoc}
\end{figure*}

\subsection{Homogeneous Ising interaction}\label{sec:homog_ising}
As a concrete example, consider the generation of a homogeneous long-range Ising interaction of the form
\begin{align}
    H_{\text{Ising}} =J \sum_{i<j}  Z_i Z_j \ ,
\label{eq:Hising}
\end{align} 
where \textit{every} pair of qubits interacts with the same strength $J$, regardless of their separation (see Fig.~\ref{fig:geometries}a). This fully connected geometry is highly nonlocal and cannot be realized through nearest-neighbor couplings alone.

The target interaction corresponds to a matrix $\boldsymbol{\varLambda}$ with elements $\boldsymbol{\varLambda}_{ij} = J$. The resulting matrix $D$ (Eq.~\eqref{eq:alpha_star}) is diagonal, such that the desired interaction can be realized using a single global driving field $f(t)$.

\subsubsection{Existing schemes}\label{sec:existing}

Before introducing optimized waveforms, it is instructive to revisit two widely used limiting cases for generating the effective Ising interaction: static~\cite{mintert2001ion} and monochromatic drivings~\cite{ospelkaus2008trapped}. These approaches offer conceptual simplicity but face trade-offs between gate speed and fidelity.

\paragraph{Static gradient}
In the case of a static magnetic field gradient, where $f(t)=1$, the accumulated geometric phases grow linearly with time as $D_l = - \nu_l T$, leading to an effective interaction matrix
\begin{align}
    \Lambda_{jk} = - T \sum_l \nu_l \eta_{jl}  \eta_{lk} \ .
\label{eq:static_ising_T}
\end{align}
This form constrains the possible interaction patterns: it naturally includes contributions from all motional modes, each weighted by its frequency $\nu_l$. While the center-of-mass (COM) mode indeed contributes equally to all qubit pairs, the presence of other modes with unequal couplings prevents the realization of a perfectly homogeneous interaction. As a result, a static global gradient alone cannot generate a fully connected Ising model with uniform couplings.

However, when combined with appropriately timed $\pi$-pulses (see Sec.~\ref{sec:spin_echo}), it becomes possible to synthesize the desired interaction geometry. One can show that at least five interaction windows ($N_I=5$ in Eq.~\eqref{eq:alphaT_sequence}) are required to approximate a homogeneous Ising interaction with vanishingly small infidelity; with fewer segments, the inhomogeneous coupling strengths cannot be compensated (see Sec.~III of the Supplemental Material~\cite{supp}).

Fig.~\ref{fig:comparison_static_monoc}(a) shows the minimum total gate duration $T$ (obtained by summing all interaction intervals while assuming instantaneous $\pi$-pulses), expressed in units of the COM mode period $2\pi/\nu_C$, as a function of the relative strength of the COM coupling $\eta_C$ (with all other mode contributions scaled accordingly). Crucially, with static gradients these are the shortest achievable times: more 
$\pi$-pulses, shorter pulses, or any other arrangement cannot lower $T$ (see Sec.~III of the Supplemental Material~\cite{supp}).

\paragraph{Modulated gradient (single–tone) drive.}
In contrast to the static case, a time-modulated magnetic-field gradient allows one to selectively enhance the contribution of a chosen motional mode $C$. The drive frequency is set close to the mode frequency, $\omega \approx \nu_C$, with detuning $\delta=\nu_C-\omega$. Under the rotating-wave approximation, the accumulated off-diagonal coupling over a gate duration 
$T = 2\pi/\delta$ is
\begin{align}
  \Lambda_{jk} = 2\pi \frac{\Omega_{jC} \Omega_{kC}}{\delta^{2}} \ ,
  \label{eq:mono_lambda}
\end{align}
where $\Omega_{jC}=\eta_{jC}\nu_C$ is the effective Rabi amplitude for ion $j$~\cite{ospelkaus2008trapped}. 

Homogeneous qubit–qubit coupling requires $\Omega_{jC}=\Omega_{kC}\equiv\Omega_C$, which can be achieved for generic ion chains only through the COM mode (or, equivalently, for the special case of \emph{two} ions using the stretch mode). Imposing a uniform target strength $J$ then fixes the detuning to
\begin{align}
  \delta = \Omega_C\,\sqrt{\frac{2\pi}{J}} \ ,
  \label{eq:mono_delta}
\end{align}
which leads to a drive frequency and gate duration
\begin{align}
  \omega = \nu_C \left(1-\sqrt{\frac{\pi}{2J}} \eta_C\right),
  \qquad
  T = \frac{2\sqrt{2\pi J}}{\eta_C \nu_C}\,.
\end{align}
Compared with the static case, the key difference is the scaling of the gate time with the qubit–motion coupling: here $T\propto \eta_C^{-1}$ (for fixed $J$ and $\nu_C$), whereas in the static case $T\propto \eta_C^{-2}$. This parametric speedup is visible in Fig.~\ref{fig:comparison_static_monoc}(a), showing up to an order-of-magnitude reduction in $T$ in the small Lamb–Dicke parameter regime.

This extra speed comes with a trade-off. Unlike the static-gradient scheme with $\pi$-pulses\,---\,which can attain essentially unit fidelity\,---\,the single-tone approach relies on approximations (\textit{e.g.}, the rotating-wave approximation and neglect of spectator modes). As shown in Fig.~\ref{fig:comparison_static_monoc}(b), these approximations are reasonably accurate in the very weak-coupling regime but lead to pronounced fidelity loss at stronger coupling, where off-resonant modes contribute more significantly. 

\begin{figure*}[t]
    \centering \includegraphics[width=0.9\textwidth]{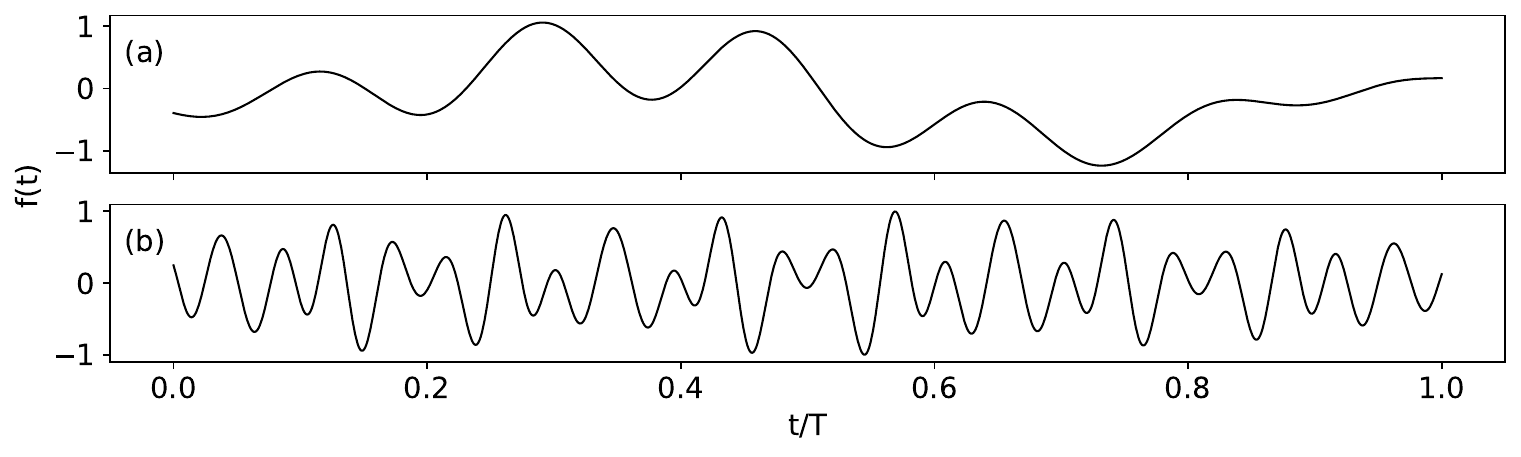}
    \caption{Global modulation $f(t)$ of the magnetic field gradient as a function of time $t$, in fractions of the total gate time $T$.  Panels (a) and (b) show distinct modulation profiles designed to implement specific quantum operations. Panel (a) shows the pulse shape used to realize a uniform Ising interaction with effective coupling $J=\pi/4$ and gate time $T=2.321/2\pi\nu$; panel (b) shows the pulse for a \textit{rainbow} coupling pattern (see Fig.~\ref{fig:geometries}(b)) with coupling $J=\pi/4$ and a longer duration $T=8.125/2\pi\nu$. The solution in (a) assumes a magnetic field gradient of $\simeq 250$~T/m and a trap frequency of $\nu/2\pi=100$~kHz, leading to a maximum qubit-motion coupling of $\eta_{j1}\simeq 0.3$. In (b), the magnetic field gradient is set to $\simeq 125$~T/m, yielding $\eta_{j1}\simeq 0.15$.}
    \label{fig:spectral}
\end{figure*}

\subsubsection{Optimized protocol}

To overcome the limitations of the schemes discussed in Sec.~\ref{sec:existing}, the framework presented here enables the direct construction of drive waveforms $f(t)$ that simultaneously satisfy the qubit-motion decoupling conditions for all motional modes and generate the desired collective qubit interaction.

As shown in Fig.~\ref{fig:comparison_static_monoc}(a), the gate time from our optimal solutions scales with the Lamb–Dicke parameter as in the single-tone case, $T \propto \eta_C^{-1}$, up to a finite constant offset (green line). Crucially, unlike the monochromatic drive, these solutions achieve vanishingly small infidelity (see Fig.~\ref{fig:comparison_static_monoc}(b)), with performance limited only by numerical precision.

Taken together, the proposed solutions combine the strengths of the existing schemes while avoiding their weaknesses: they retain the speed of the monochromatic approach and the high-fidelity of the static case. 

This capability is especially valuable in the strong-coupling regime\,---\,precisely where conventional protocols are least effective\,---\,yet where fast gate operations become possible (see Fig.~\ref{fig:comparison_static_monoc}). For a concrete illustration, consider a large magnetic-field gradient yielding a COM-mode coupling of $\eta_{C}=0.3$. Such operating points are consistent with a maximum gradient of $\partial B/\partial z \simeq 250~\mathrm{T/m}$ and a trap frequency of $\nu/2\pi \simeq 100~\mathrm{kHz}$, values within current experimental reach~\cite{gerasin2025surface,weidt2016trapped}.

\begin{figure*}[t]
    \centering \includegraphics[width=0.99\textwidth]{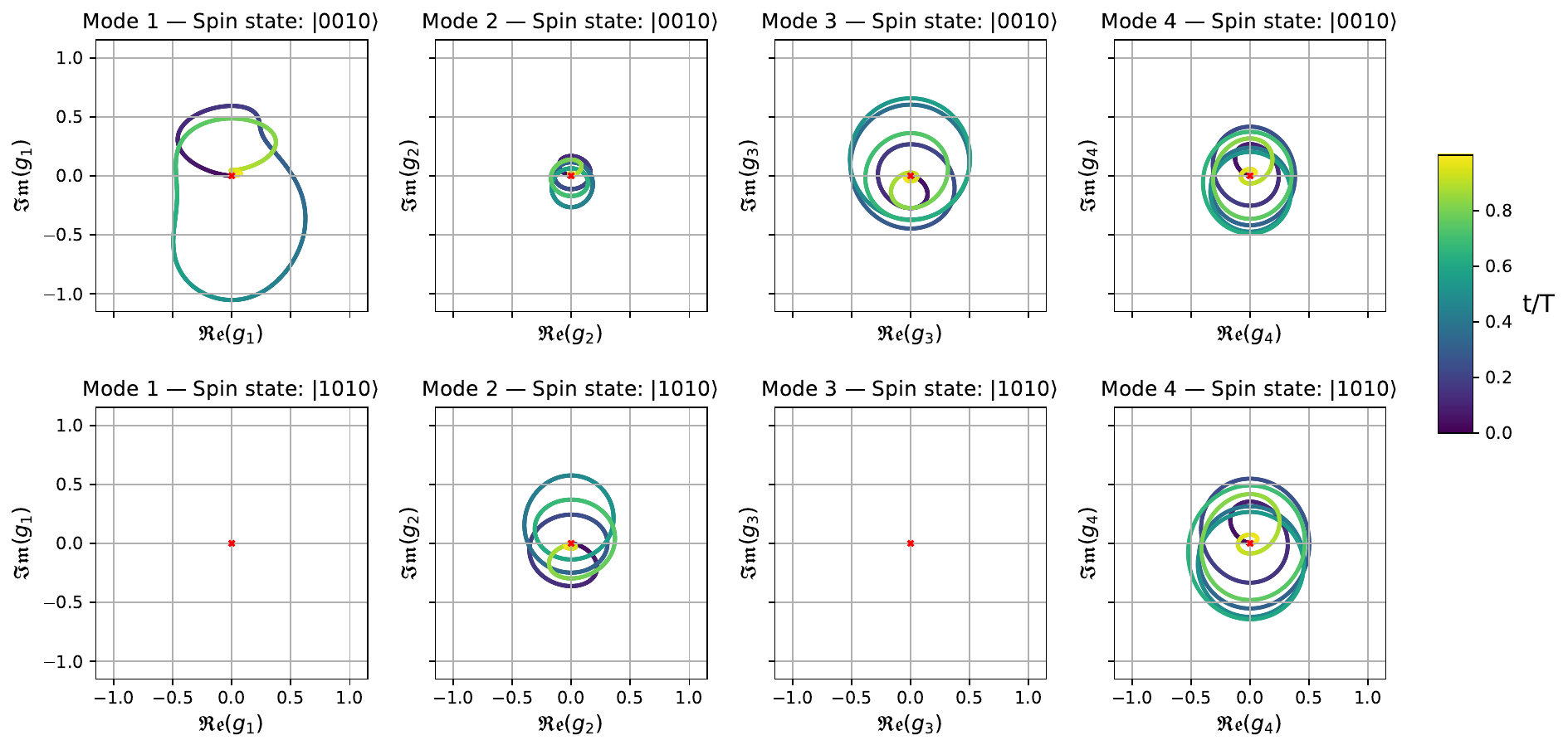}
    \caption{Phase-space trajectories of the four motional modes in a four-ion chain during an effective Ising gate with target strength $J=\pi/4$. Columns correspond to $l=1,\dots,4$ (left to right). In each panel, the real and imaginary parts of the displacement $g_l(t)$ are shown, with time indicated by color from dark blue (start) to yellow (end); the red cross marks the displacement at the gate time $T$. Row (a) shows the evolution for the initial qubit state $\ket{0010}$, and row (b) for $\ket{1010}$, highlighting the qubit-dependent nature of the dynamics.}
    \label{fig:dynamic_alpha_i}
\end{figure*}

For a target interaction strength of $J=\pi/4$ (see Eq.~\eqref{eq:Hising}), the optimal drive depicted in Fig.~\ref{fig:spectral}(a) allows the gate to be completed in just two oscillation cycles of the COM mode\footnote{In this example, we use the boundary conditions that match the monochromatic case; but analogous results hold for alternative boundary choices.}.
This corresponds to a total gate time of $T \simeq 2.3 \cdot (2\pi/\nu_1)\simeq 23~\mu s$, which is comparable to the fastest multi-qubit gate implementations reported in trapped-ion systems~\cite{bruzewicz2019trapped}.
The resulting gate infidelity is limited only by the numerical precision of the pulse shaping (set to $10^{-9}$), and exceeds the accuracy typically achieved by existing multi-qubit gate protocols by several orders of magnitude.

A key advantage of the present approach is that it eliminates the need to spectrally resolve individual sideband transitions associated with specific motional modes. Instead, the gate is driven through the simultaneous excitation of all collective modes, with all resonant and off-resonant processes incorporated exactly in the evolution. This capability is illustrated in Fig.~\ref{fig:dynamic_alpha_i}, which shows the phase-space trajectories of the four motional modes during the execution of the gate.

We consider initial motional states in the ground state $\ket{0}^{\otimes 4}$, and depict the evolution of each mode in a separate panel.  To highlight the qubit-dependent nature of the dynamics, we consider two distinct initial qubit configurations: $\ket{0010}$ and $\ket{1010}$, corresponding to the first and second rows, respectively. The trajectory of each mode in phase space is shown using a color gradient from blue to yellow to indicate the time evolution, and the final point at gate time is marked with a red cross for clarity.

In the case of the qubit state $\ket{0010}$, all four motional modes are displaced during the gate, undergo closed trajectories in phase space, and return precisely to the origin at the final time $T$. This ensures full qubit–motion decoupling and corresponds to a unitary evolution purely within the qubit subspace. The total geometric phase accumulated is proportional to the net area enclosed by the trajectories, summed over all modes. Notably, all four modes contribute nontrivially and constructively to the generation of the desired entangling operation.

For the alternative initial qubit state $\ket{1010}$, the symmetry properties of the coefficients $\eta_{jl}$ result in only two of the modes being excited, while the other two remain stationary at the origin throughout the evolution. This behavior highlights a key feature of the protocol: qubit–motion entanglement and its cancellation are achieved without the need for individual mode addressing. Instead, the collective and state-dependent structure of the drive naturally determines which modes participate. As in the previous case, the excited modes follow closed trajectories and return to the origin at time $T$, ensuring complete qubit–motion decoupling.

Fig.~\ref{fig:dynamic_alpha_i} illustrates the trajectories for a motional ground‐state $\ket{0}^{\otimes 4}$; the same closed‐loop behavior holds even when each mode begins in a different Fock state, or even in a thermal mixture. In fact, since the displacements $\Delta_l(t)$, and thus motion, depend only on the drive $f(t)$ and the qubit configuration\,---\,not on the initial phonon number\,---\,any initial occupation $n_l(0)$ simply adds a constant offset to the total phonon count without altering the trajectory shape. Concretely, the phonon‐number evolution is
\begin{align}
    \langle a_l^{\dagger}(t)a_l(t) \rangle = n_l(0) + \abs{\Delta_l(t)}^2 \langle S_l^2 \rangle_{\text{qubit}} \ ,
\label{eq:n_intime}
\end{align}
where $\langle \cdot \rangle$, $\langle \cdot \rangle_{\text{qubit}}$ denote expectation values over the full system and qubit subsystem respectively.  In a thermal state, the term $\abs{\Delta_l(t)}^2 \langle S_l^2 \rangle_{\text{qubit}}$ is added identically to each component. Since the protocol enforces $\Delta_l(T)=0$ (so that every phase‐space trajectory closes), each state $\ket{n_l(t)}$ returns to exactly $n_l(0)$. As a result, all qubit‐motion entanglement vanishes at $t=T$ regardless of the initial temperature, and no ground‐state cooling is required. Crucially, this property holds for any gate constructed within our framework and is not specific to the example shown here.

\subsection{Rainbow interaction}
\label{sec:rainbow}

Another interesting application of the present framework is the ability to engineer a global entangling gate of the form 
\begin{align}
U=\exp(i H_{\text{R}}) \ , \quad  H_{\text{R}} = J\sum_{k} Z_k Z_{N-k+1} \ ,
\label{eq:Urainbow}
\end{align}
where $J$ sets the pairwise interaction strength between symmetrically located qubits. This gate can be realized using a single global drive $f(t)$ engineered to induce $ZZ$-type interactions across reflection-symmetric qubit pairs. When combined with appropriate single-qubit rotations (implemented via standard techniques; see Sec.~V of the Supplemental Material~\cite{supp}), it enables the deterministic preparation of a \textit{rainbow state}~\cite{ramirez2015entanglement}
\begin{align}
    \ket{\Psi_{\text{rainbow}}} = \bigotimes_{j=1}^{N/2} \ket{\psi^-}_{\,j ,\,N-j+1} \ ,
\label{eq:rainbow_state}
\end{align}
where $\ket{\psi^{-}}_{a,b} = \frac{1}{\sqrt 2} \left( \ket{0}_a\ket{1}_b - \ket{1}_a\ket{0}_b \right)$ is the singlet state between sites $a$ and $b$. The resulting state, illustrated in Fig.~\ref{fig:geometries}b, consists of qubits paired symmetrically across the chain’s center, forming a structured network of long-range singlets. Notably, rainbow states exhibit volume law scaling of entanglement entropy~\cite{vitagliano2010volume}, where the entropy grows linearly with subsystem size. This behavior stands in sharp contrast to the area law observed in the ground states of local Hamiltonians, making rainbow states a powerful tool for investigating quantum many-body entanglement and non-equilibrium dynamics~\cite{hastings2007area, ramirez2015entanglement, langlett2022rainbow}. 

To illustrate the approach, a four-qubit rainbow gate is implemented with $J=\pi/4$. The optimized driving profile $f(t)$ is shown in Fig.~\ref{fig:spectral}(b), assuming a Lamb–Dicke parameter $\eta_{C} = 0.15$ for the COM mode, with all other mode couplings adjusted consistently. Due to the weaker qubit–phonon interaction and the constraints imposed by the desired coupling geometry, the required gate duration is slightly longer, with $T\simeq 8.1\cdot(2\pi/\nu_C)\simeq 81 ~\mu s$. 

\begin{figure}[t]
    \centering \includegraphics[width=0.49\textwidth]{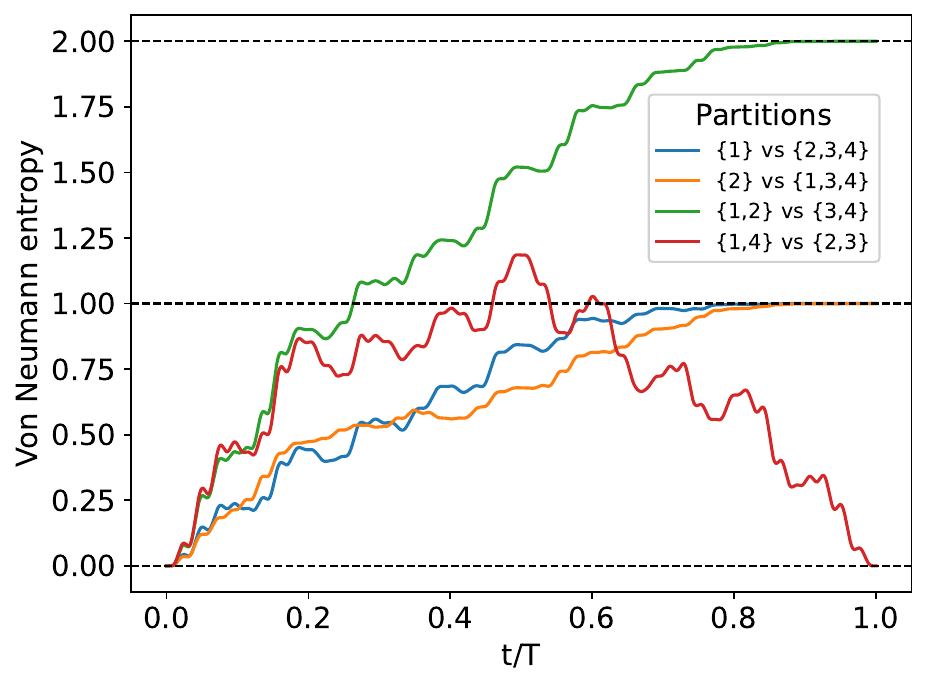}
    \caption{Dynamics of the von Neumann entropy $S$ for various qubit bipartitions in a four-ion system, shown as a function of the normalized gate time $t/T$. The results correspond to the application of an optimized global drive $f(t)$ designed to generate the gate in Eq.~\eqref{eq:Urainbow}.}
    \label{fig:entropies}
\end{figure}

The dynamics of entanglement generation under this gate are illustrated in Fig.~\ref{fig:entropies}, which displays the evolution of the von Neumann entropy  $S=-\Tr(\rho\log \rho)$ for several qubit bipartitions throughout the gate operation. Each curve corresponds to a specific partition of the qubit register, either single-qubit subsystems (\textit{e.g.} qubit $\{1\}$ or $\{2\}$) or two-qubit subsets (\textit{e.g.} qubits $\{1,2\}$ or $\{1,4\}$). Time is normalized by the total gate duration $T$, and the system is initialized in the separable state  $\ket{+}^{\otimes 4} \otimes \ket{0}^{\otimes 4}$, with qubits in superposition and motional modes in the ground-states.  

As the global drive $f(t)$ induces phonon-mediated qubit–qubit interactions, entanglement builds up across the system in a reflection-symmetric pattern. The von Neumann entropy of qubit $1$ increases during the protocol and saturates at $S=1$, indicating that it becomes maximally entangled with the rest of the system.  A similar trend is observed for qubit $2$, although the temporal profile differs due to the specific values of its qubit–phonon coupling constants $\eta_{jl}$. The entropy of the pair $\{1,2\}$ reaches $S=2$, reflecting maximal entanglement of both qubits with the remainder of the system. 

The bipartition $\{1,4\}$, which is expected to form a Bell pair in the final rainbow state, exhibits a notably non-monotonic entropy evolution. In the first half of the protocol, the entropy of this pair rises above $S=1$, indicating that qubits $1$ and $4$ become entangled not only with each other but also with the rest of the system, participating in more complex multiqubit correlations. In the second half of the gate evolution, however, the entropy steadily decreases and ultimately vanishes, confirming that qubits $1$ and $4$ become disentangled from the remainder of the register and form a pure, two-qubit state.

\begin{figure*}[t]
    \centering \includegraphics[width=0.9\textwidth]{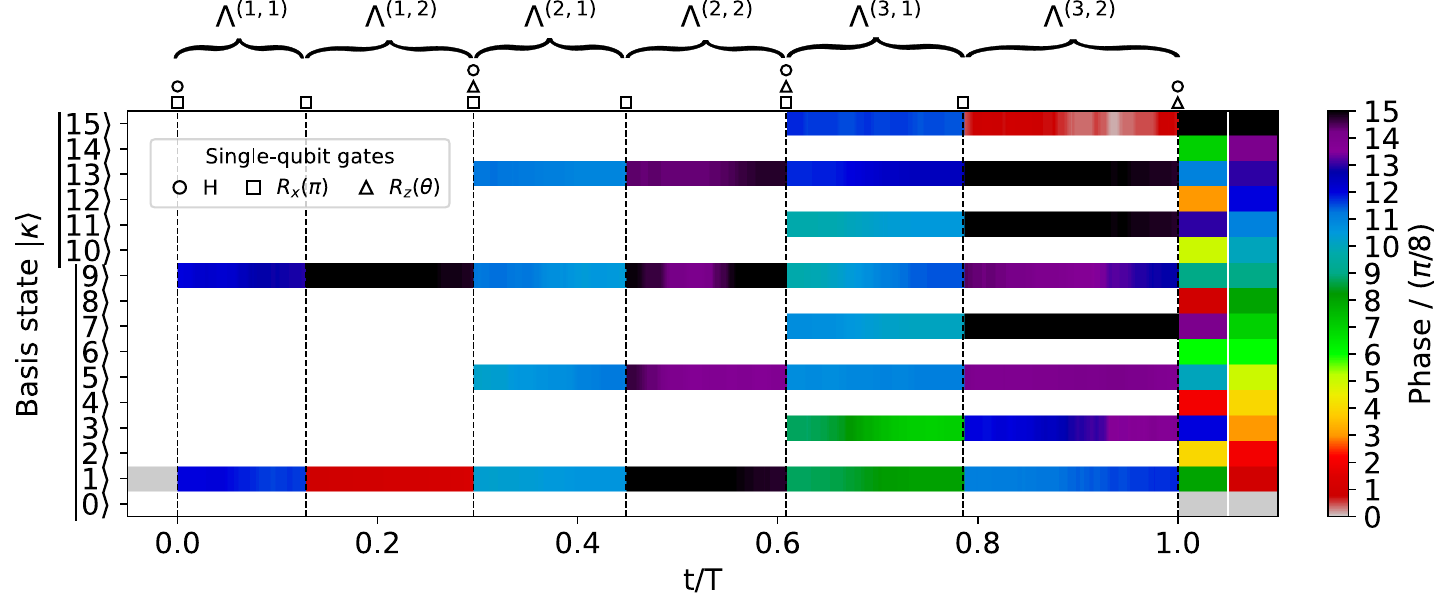}
    \caption{Phase dynamics of levels with finite occupation during the implementation of the four-qubit quantum Fourier transform (QFT), simulated for the input state $\ket{1}\equiv\ket{0001}$. The horizontal axis shows the normalized time $t/T$, the vertical axis labels computational basis states $\ket{\kappa}$, and the color encodes the instantaneous phase of each component in units of $\pi/8$. Basis states whose amplitude remains negligible throughout the evolution are shown in white, allowing phase-zero regions of unoccupied states to be distinguished from genuine phase-zero values of populated components. Vertical dashed black lines mark the instants at which single-qubit gates are applied; different marker shapes above each line indicate Hadamard, $R_x(\pi)$, and $R_z(\theta)$ rotations on the qubits specified by the circuit in Fig.~S1. The intervals between gate layers correspond to entangling evolution under the optimized global magnetic-field gradients $f_{\ell,p}(t)$, which implement the elementary interaction matrices $\Lambda^{(\ell,p)}$. Within each stage-$\ell$, the combined effect of these segments yields the effective interaction matrix $\boldsymbol{\varLambda}^{(\ell)}$ defined in Eq.~\eqref{eq:Lambda-stage-k}. Just before the final local layer, the phase pattern still appears to differ from the QFT target, but it is already in the correct form up to the concluding Hadamard and $Z$ rotations, which produce the exact four-qubit QFT spectrum with phases $\exp(i \kappa \pi/8)$. The final SWAP layer (solid white line), which can be implemented using the protocol of Sec.~\ref{sec:rainbow}, simply reorders the outputs to obtain the standard QFT ordering. A small padding before $t=0$ and after $t=T$ makes the initial and final phases visually explicit and separates the coherent evolution from the final SWAPs.}
    \label{fig:QFT_from_1}
\end{figure*}

\subsection{Quantum Fourier transform}
\label{sec:qff}

The preceding examples demonstrate how carefully tailored \textit{global} magnetic field gradients can produce specific qubit-qubit interaction patterns. 
Many unitaries, such as the \textit{quantum Fourier transform} (QFT), however, require several steps of dynamics controlled by a magnetic field gradient interleaved with single-qubit gates.

The QFT is realized by a sequence of Hadamard gates $\frac{1}{\sqrt{2}}(X+Z)$, a series of controlled phase gates after every single Hadamard gate,
and a set of SWAP gates at the end of the circuit. The dynamics of all the SWAP gates reads $\exp\left(i\frac{\pi}{4}H_S\right)$ with $H_S=\sum_{j}X_jX_{N-j+1}+Y_jY_{N-j+1}+Z_jZ_{N-j+1}$.
Since all the terms in the exponent commute, this can be realized as a sequence of unitaries induced by $\sum_{j}X_jX_{N-j+1}$, by $\sum_jY_jY_{N-j+1}$ and by $\sum_jZ_jZ_{N-j+1}$ respectively.
Together with single-qubit unitaries, this is the dynamics discussed above in Sec.~\ref{sec:rainbow}.
In the following, we will thus take for granted that the SWAP operations can be implemented and focus on the remaining part of the quantum Fourier transform centered around controlled phase gates.

Even though all the controlled phase gates applied after any Hadamard gate commute and can thus be realized simultaneously, most hardware platforms require a sequential realization.
The present framework, however, permits a simultaneous realization in a rather natural fashion. 

Recall that a controlled phase
\begin{align}
\mathrm{CP}(\varphi)=\exp(i\varphi\,\dyad{1}\!\otimes\!\dyad{1})
\end{align}
differs from a $ZZ$ interaction
\begin{align}
\exp(i\theta\, Z\!\otimes\! Z)
\end{align}
only by single–qubit $Z$ rotations and an overall phase, because of the identity
\begin{align}
\dyad{1}\!\otimes\!\dyad{1}
=\frac{1}{4} \!\left(\mathbb 1\!\otimes\! \mathbb 1 - Z\!\otimes\! \mathbb 1 - \mathbb 1\!\otimes\! Z + Z\!\otimes\! Z\right) \ .
\end{align}
Thus, implementing the QFT reduces to realizing a pattern of $ZZ$ couplings with angles $\theta=\varphi/4$; and the accompanying single–qubit $Z$ phases can be folded into the surrounding Hadamards.

With the convention that the QFT applies a Hadamard on qubit $k$ and then controlled phases from qubit $k$ to all \emph{later} qubits $j\in\{k{+}1,\dots,N\}$, the required controlled-phase angles are
\begin{align}
\varphi_{j k} = \frac{\pi}{2^{\,j-k}}\qquad (j>k) \ .
\label{eq:qft-cphase}
\end{align}
Equivalently, the corresponding $ZZ$ phases in \emph{stage–$\ell$} define a target interaction matrix with entries
\begin{equation}
\boldsymbol{\varLambda}^{(\ell )}_{jk} =
\begin{cases}
\dfrac{\pi}{2^{|j-k|+2}} \ , & \ell = \min(j,k),\ \max(j,k) > \ell \ , \\
0 \ , & \text{otherwise} \ .
\end{cases}
\label{eq:Lambda-stage-k}
\end{equation}
For a four–qubit example, stages $\ell \in \{1,2,3\}$ are required.
In each case, the target $\boldsymbol{\varLambda}^{(\ell)}$ can be synthesized as a linear combination of two directly realizable interaction matrices $\Lambda^{(\ell, p)}$ according to Eq.~\eqref{eq:alphaT_sequence} (\textit{i.e.} $\boldsymbol{\varLambda}^{(\ell)} = O^{(1)}\Lambda^{(\ell, 1)}\left(O^{(1)}\right)^{\top} + O^{(2)}\Lambda^{(\ell, 2)}\left(O^{(2)}\right)^{\top}$), enabling all controlled phases $\Lambda^{(\ell, p)}$ to be effected \emph{simultaneously}. The full circuit is shown in Fig.~S1 of the Supplemental Material~\cite{supp}.

We validate the construction by simulating the full time evolution\,---\,including all additional gates in the circuit\,---\,using tailored drive profiles $f(t)$ that, at each stage, generate the effective couplings of Eq.~\eqref{eq:Lambda-stage-k}. Fig.~\ref{fig:QFT_from_1} shows the resulting implementation of the QFT on a four-qubit register initialized in $\ket{1}\equiv\ket{0001}$.
The vertical axis displays the computational basis states $\{\ket{\kappa}\}_{\kappa=0}^{15}$, while the horizontal axis shows the normalized time $t/T$.
The color scale encodes the time-dependent phase of each basis state with finite occupation.

Long bars correspond to intervals of entangling evolution under the
optimized global fields $f_{\ell, p}(t)$ in each stage-$\ell$, which generate the effective
couplings $\Lambda^{(\ell, p)}$. The vertical dashed black lines mark the instants at
which additional local gates are applied. These local operations (Hadamard gates, $\pi$-pulses and $Z$ rotations) are treated as
instantaneous on the timescale of the entangling dynamics and are indicated by different marker shapes (circle, square, triangle) on top of each
dashed line. 

The protocol starts with the state in $\ket{1}$, after which a Hadamard and a $\pi$-pulse redistribute amplitude and phase between $\ket{1}$ and $\ket{9}$. The first entangling window, driven by the optimized profile $f_{1,1}(t)$, then implements the effective coupling $\Lambda^{(1,1)}$, slightly reshaping the phase landscape. At $t/T \simeq 0.16$ a further $\pi$-pulse is applied, which now produces a clear phase contrast between $\ket{1}$ and $\ket{9}$; this is followed by a second entangling window that realizes $\Lambda^{(1,2)} = -\Lambda^{(1,1)}$, and a final $\pi$-pulse. Importantly, the drive profiles associated with each matrix $+\Lambda^{(\ell, p)}$ and $-\Lambda^{(\ell, p)}$ are obtained from independent optimizations. Knowing a solution $f_{\ell, p}(t)$ that realizes $+\Lambda^{(\ell, p)}$ does not, in general, determine a corresponding transformation of $f_{\ell, p}(t)$ that would produce $-\Lambda^{(\ell, p)}$; this is reflected in the bars of different length $T_{\ell, p}/T$.

The same pattern repeats in the subsequent stages. Each Hadamard increases the number of computational-basis components that carry population, the following entangling interval imprints the appropriate collective phase pattern on this enlarged set, and intermediate single-qubit $Z$ rotations modify relative phases on individual basis states. At the end of the protocol, a final Hadamard gate together with two SWAP gates (indicated by the solid white line), which restore the standard QFT ordering in $\kappa$, produce the desired output state
\begin{align}
    \mathrm{QFT}\,\ket{1}
    = \frac{1}{\sqrt{2^N}} \sum_{\kappa=0}^{2^N-1} e^{\frac{2\pi i \kappa}{2^{N}}} \ket{\kappa} \ .
\end{align}
For $N=4$, this corresponds to a uniform superposition with amplitudes $1/4$ and phases $e^{i \pi \kappa/8}$ across all basis states $\ket{\kappa}$.

Taken together, these results show that the engineered global-control sequence reproduces ideal QFT dynamics. The crucial advantage is \emph{simultaneity}: all controlled phases associated with a given control qubit are generated in one shot. For large registers this replaces a long sequence of pairwise gates (scaling with qubit number $N$) by a small number of global entangling windows (one per stage), reducing depth, easing scheduling, and improving scalability by concentrating the interaction into fewer, more coherent operations.

\begin{figure*}[t]
    \centering \includegraphics[width=0.8\textwidth]{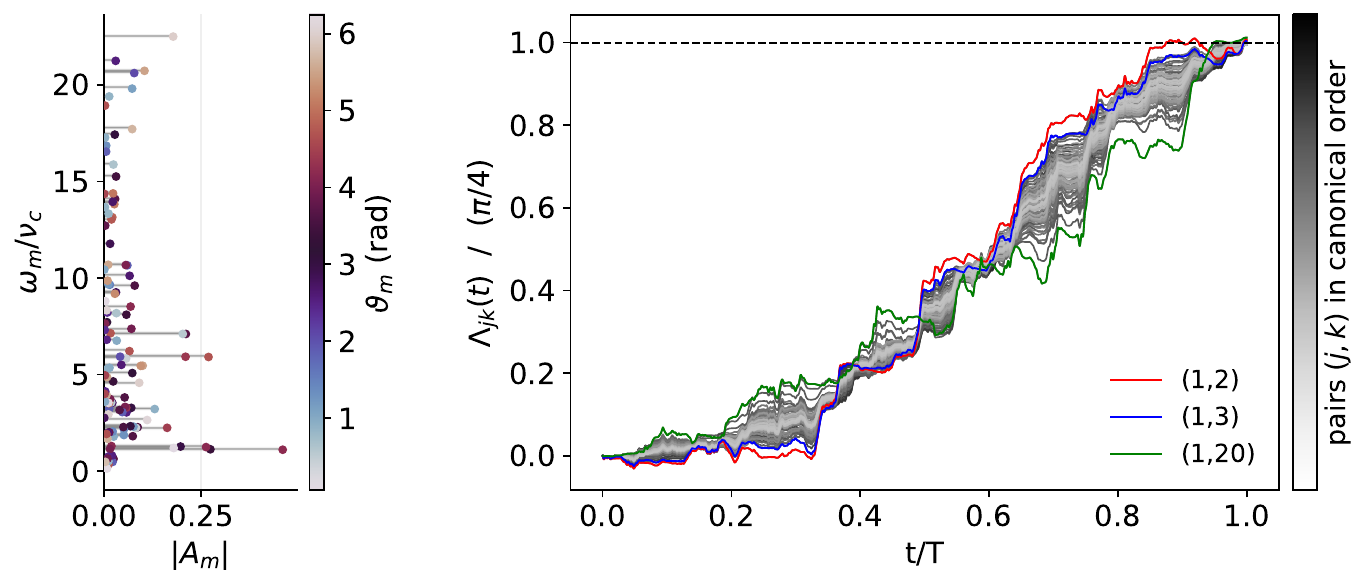}
    \caption{\textbf{Left.} Frequency--phase spectrum of the global drive
    $f(t)=\sum_{m} A_{m}\cos(\omega_{m} t+\vartheta_{m})$ used to implement a
    homogeneous Ising interaction with target pairwise phase $J=\pi/4$   on a chain
    of $N=20$ ions (with $\eta = 0.2$) in a total time $T \simeq 8.95\times(2\pi/\nu_c)$. Each stick at $\omega_{m}$ has height $|A_{m}|$ and marker
    color encodes the phase $\vartheta_{m}$. \textbf{Right.} Time evolution of the implemented pairwise phases $\Lambda_{jk}(t) = \int_0^t dt' \alpha(t')_{jk}$ as a function of normalized time $t/T$ for the same $N=20$ ion chain. Curves are shown normalized by the
    target ($\Lambda_{jk}(t)/J$), so the horizontal dashed line at unity marks
    the desired uniform coupling. A few representative pairs $(1,2)$, $(1,3)$,
    and $(1,20)$ are highlighted in red, blue and green. By the gate time $t=T$, all off-diagonal phases collapse onto the
    target line, certifying that the homogeneous interaction pattern is achieved. At gate time, motional displacements are closed for every mode, and the simulated gate infidelity is below $10^{-4}$.}
    \label{fig:phases_20_ions}
\end{figure*}

\section{Quantum gates for larger qubit registers}
\label{sec:largesystem}

A key advantage of the present framework is that gate synthesis does not require simulating the full system dynamics. Instead, the driving schemes are constructed directly from the harmonic dynamics of the individual modes. As a result, the method remains applicable even for systems where a full simulation of the joint qubit–motional evolution is infeasible due to the exponential growth of the Hilbert space.

Although the driving schemes themselves can be designed efficiently, a brute–force, state–vector–based evaluation of the resulting gate fidelity would still incur an exponential cost. This bottleneck can be avoided by using an efficiently computable lower bound on the worst–case gate fidelity, as we now describe.

\subsection{Bound on gate fidelity}

Within the present framework, an entangling gate is specified as $U= \exp(-i\sum_{j\neq k} \Lambda_{jk} Z_j Z_k)$ in terms of a real symmetric matrix $\Lambda$ with vanishing diagonal.
Let $\boldsymbol{\varLambda}$ denote
the target coupling matrix and $\Lambda$ the one actually realized, and define the deviation $\delta\Lambda = \Lambda - \boldsymbol{\varLambda}$. 

As shown in Sec.~VI of the Supplemental Material~\cite{supp}, the worst–case gate fidelity obeys the bound
\begin{align}
F_{\min}
\;\ge\; \cos^2\!\bigl(N\,\|\delta\Lambda\|_2\bigr) \ ,
\label{eq:fmin}
\end{align}
valid whenever $N\,\|\delta\Lambda\|_2 \le \pi/2$. In this regime the bound is
faithful (it approaches unity as $\delta\Lambda\to 0$) and decreases
monotonically with increasing error. Crucially, the operator $2$-norm
$\|\delta\Lambda\|_2$ can be evaluated with an effort polynomial in $N$, so
Eq.~\eqref{eq:fmin} provides an efficiently computable and scalable certificate of gate performance, suitable as a post-optimization quality guarantee.

\subsection{Homogeneous Ising interaction on a $20$-qubit register}

Gate synthesis for larger qubit registers and the assessment of gate fidelities in terms the bound Eq.~\eqref{eq:fmin} is exemplified in the following with the target of a uniform all-to-all Ising interaction (see Sec.~\ref{sec:homog_ising}) and a chain of $N=20$ ions.

A representative solution is shown in Fig.~\ref{fig:phases_20_ions} for an optimized solution with $6N$ basis elements in the expansion $f(t)=\sum_{m}A_{m}\cos(\omega_{m}t+\vartheta_{m})$. The left panel displays the spectrum of the optimal global control field, where the vertical stick marks each
frequency component $\omega_m$ (plotted in units of the trap frequency $\nu_c$), its height encodes the relative amplitude $|A_m|$, and its color represents the phase $\vartheta_m$.
The spectrum is dominated by tones within a few multiples of the lowest motional frequency $\nu_c$, even though the optimization also makes use of a sparse tail of higher-frequency components extending up to $\simeq 20\nu_c$.
These higher tones primarily address the upper normal modes of the chain, whose frequencies reach up to $\simeq 11\nu_c$, while contributing only weakly to the dynamics.

Evaluating the norm-based bound on the worst–case gate fidelity,
Eq.~\eqref{eq:fmin}, yields $F_{\mathrm{min}}>0.96$, so the protocol is
certified to operate in the high-fidelity regime even without access to the
full state dynamics. In this $N=20$ example, a direct evaluation of the gate performance is still feasible and shows that the actual average fidelity is substantially higher: computing the process fidelity $F_{\mathrm{pro}} = \bigl|\mathrm{Tr}\big(U_{\mathrm{r}}^{\dagger} U_{\mathrm{T}}\big)\bigr|^2 / d^2$ and the corresponding average gate fidelity $F_{\mathrm{avg}} = (d F_{\mathrm{pro}}+1)/ (d+1)$, with $d=2^N$, yields an average gate infidelity $1 - F_{\mathrm{avg}} \lesssim 10^{-4}$. This confirms that the norm-based
bound is conservative yet informative for the parameters of interest.

The right panel of Fig.~\ref{fig:phases_20_ions} illustrates the corresponding time evolution of the implemented pairwise phases $\Lambda_{jk}(t)=\int_{0}^{t}dt'\,\alpha_{jk}(t') $ plotted versus the normalized time $t/T$ for every pair $(j,k)$ in the $20$-ion chain. Each trajectory is normalized by the target interaction strength $J$, so that the horizontal dashed line at unity marks the ideal homogeneous Ising phase. Grey curves show the phases for all pairs $(j,k)$, ordered according to the color gradient, while three representative cases\,---\,$(1,2)$, $(1,3)$, and the long-range pair $(1,20)$\,---\,are highlighted. Pairs involving neighboring ions, such as $(1,2)$ and $(1,3)$, have very similar phase trajectories $\Lambda_{12}(t)$ and $\Lambda_{13}(t)$, in clear contrast to that of $\Lambda_{1,20}(t)$ for the distant pair $(1,20)$. This is consistent with the fact that, in larger ion chains, the qubit–motion couplings $\eta_{jl}$ are very similar for nearby ions, while ions that are far apart exhibit more distinct participation patterns. Despite these microscopic variations, all $\Lambda_{jk}(t)$ curves converge cleanly to the target value $J$ at $t=T$, certifying that the gate implements a uniform Ising interaction across the entire chain.

\section{Conclusions}

As the growth of registers of engineered qubits creates expectations to use quantum devices for practical computations, it becomes increasingly important to realize high-fidelity gates in such large registers.
Careful gate design has resulted in impressive improvement of gate fidelities in devices with few (often two) qubits. Translating these achievements into the large registers that are necessary for practical applications is a highly non-trivial task.

The present approach achieves this in terms of an effective inter-qubit interaction that is mediated by all motional modes simultaneously instead of the common technique to spectrally select one single mode.
Since the underlying optimal control problem does not require any numerical simulation, this approach is not limited by common restrictions to system size of few qubits.

As exemplified with the quantum Fourier transform, the present approach is also very well suited for gate parallelization.
The ability to apply mutually commuting gates in parallel, has the potential to improve the scaling behavior of quantum algorithms when fully scalable hardware is available.
In the current era of devices with limited coherence time, the ability to perform quantum gates in parallel can mean that an algorithm can be executed within the restrictions of finite coherence time, even if a fully sequential implementation would exceed these restrictions.

The resultant ability to implement multi-qubit algorithms both faster and with higher accuracy offers a concrete route to extracting more value from today’s imperfect hardware, while simultaneously laying the groundwork for the larger, fault-tolerant devices envisioned for the future.

\section{Acknowledgements}

We gratefully acknowledge stimulating discussions with Nguyen Le and Jungsang Kim, 
and the use of the Imperial College London Research Computing Service 
(DOI: 10.14469/hpc/2232). This work was supported by the U.K. Engineering and 
Physical Sciences Research Council through the EPSRC Hub in Quantum Computing 
and Simulation (EP/T001062/1).

\bibliography{biblio}

\clearpage
\onecolumngrid 
\appendix
\section*{Supplemental Material}

\beginsupplement

\section{Time-dependent polaron transformation} \label{appendixa}

This appendix collects the analytical expressions required to evaluate and optimize the dynamics under a time-dependent magnetic gradient $f(t)$. Closed-form formulas are provided for the mode displacements, the boundary conditions on the global drive, and the accumulated phases $D_l$ that determine the effective qubit–qubit couplings, in a form directly usable for the optimization of a target interaction matrix $\boldsymbol{\varLambda}$.

The time-dependent polaron transformation Eq.~(3) of the main text relies on the time-dependent functions $g_l(t)$, which are solutions to the equations of motion specified in Eq.~(4) of the main text. These functions are given by

\begin{align}
    g_l(t) = e^{-i t \nu_l} \left( g_l(0) + \nu_l \int_0^t d\tau e^{i \nu_l \tau } f(\tau) \right) \ ,
    \label{eq:g_l}
\end{align}
where $f(t)$ denotes the global field, and $\nu_l$ represents the motional frequency of mode $l$. To ensure consistency between the transformed and original frames, the functions $g_l(t)$ must satisfy the appropriate boundary conditions, as explained in Sec.~II C of the main text. 

Considering the magnetic field gradient parametrization in Eq.~(20) of the main text, the boundary conditions that are consistent with those of the static gradient (Sec.~II C 1 of the main text) can be expressed as the following system of equations
\begin{subequations}
\begin{align}
0&=\sum_j \frac{A_j\omega_j}{\omega_j^2-\nu_l^2}\Big[
 \nu_l  \sin(\nu_l T) \sin(\vartheta_j)  + \omega_j \cos(\omega_j T + \vartheta_j)  - \omega_j \cos(\nu_l T) \cos(\vartheta_j)
\Big] \\
0&= \sum_j \frac{A_j\omega_j}{\omega_j^2-\nu_l^2}\Big[
 \nu_l \cos(\nu_l T) \sin(\vartheta_j)  - \nu_l \sin(\omega_j T + \vartheta_j)  + \omega_j \sin(\nu_l T) \cos(\vartheta_j)
\Big] \ , 
\end{align}
\label{eq:bc_constraints}
\end{subequations}
in terms of the motional frequencies $\nu_l$, the gate time $T$, and the spectral components of the drivings $(A_j, \omega_j, \vartheta_j)$. 

Similarly, the boundary conditions that are consistent with those of a monochromatic oscillating magnetic field gradient (see Sec.~II C 2 of the main text) correspond to the system of equations
\begin{subequations}
\begin{align}
    0&=\sum_j \frac{A_j}{\omega_j^2 - \nu_l^2} \Big[-\omega_j \sin(\vartheta_j)  + \omega_j \cos(\nu_l T)\sin(\omega_j T + \vartheta_j)  - \nu_l \sin(\nu_l T) \cos(\omega_j T + \vartheta_j) \Big]\\
    0 &= \sum_j \frac{A_j}{\omega_j^2 - \nu_l^2} \Big[ - \nu_l\cos(\vartheta_j)  + \nu_l\cos(\nu_l T)\cos( \omega_j T +\vartheta_j)  + \omega_j\sin(\nu_l T) \sin( \omega_jT +\vartheta_j) \Big] \ . 
\end{align}
\label{eq:bc_constraints_oscill}
\end{subequations}
Since Eqs.~\eqref{eq:bc_constraints} or Eqs.~\eqref{eq:bc_constraints_oscill} must be satisfied for each motional mode $l$, a parametrization using $2N$ sets $(A_j, \omega_j, \vartheta_j)$\,---\,for a system with $N$ motional modes\,---\,can automatically fulfill the conditions. Typically, while $2N$ sets of parameters are sufficient, adding at least one extra pair is required for tailoring the dynamics as desired. 

For a field of the form given in Eq.~(20) of the main text, the explicit expressions for the accumulated phases $D_l$ that determine the interaction matrices $\Lambda$ can be derived. To facilitate this, the imaginary part of $g_l(t)$ is first determined to be
\begin{align}
\mathfrak{Im} (g_l) &= \sum_j \frac{A_j}{\omega_j^2 - \nu_l^2} \Bigl[ \nu_l^2 \cos(\omega_j t + \vartheta_j)  - \omega_j^2 \cos(\nu_l t) \cos(\vartheta_j)  + \omega_j \nu_l \sin(\nu_l t) \sin(\vartheta_j) \Bigr] \ .
\end{align}
This leads to the derivation of the explicit time-dependent phases for each motional mode $l$
\begin{align}
\frac{\Phi_l(t)}{\nu_l}
= \sum_{j,k} \frac{A_j A_k}{\omega_j^2 - \nu_l^2}\cos(\omega_k t+\vartheta_k)
\Bigl[&\nu_l^2 \cos(\omega_j t+\vartheta_j)  - \omega_j^2 \cos(\vartheta_j) \cos(\nu_l t)  + \omega_j \nu_l \sin(\vartheta_j) \sin(\nu_l t)\Bigr]
\end{align}
and, then, the accumulated phases $D_l=\int_0^T \Phi_l(t)$ can be expressed in the closed form
\begin{widetext}
\begin{align}
\frac{D_l}{\nu_l}
&= \sum_{j,k}\frac{A_jA_k}{2(\omega_j^2 - \nu_l^2)}\Bigg\{
\;2\nu_l^2\Bigg[
\frac{\sin\!\Big(\tfrac{(\omega_k-\omega_j)T}{2}\Big)}{\omega_k-\omega_j}\,
\cos\!\Big(\vartheta_k-\vartheta_j+\tfrac{(\omega_k-\omega_j)T}{2}\Big)
+
\frac{\sin\!\Big(\tfrac{(\omega_k+\omega_j)T}{2}\Big)}{\omega_k+\omega_j}\,
\cos\!\Big(\vartheta_k+\vartheta_j+\tfrac{(\omega_k+\omega_j)T}{2}\Big)
\Bigg] \nonumber\\[0.25em]
&\qquad
-\omega_j^2\cos\vartheta_j\Bigg[
\frac{\sin\!\big((\nu_l+\omega_k)T+\vartheta_k\big)-\sin\vartheta_k}{\nu_l+\omega_k}
+
\frac{\sin\!\big((\nu_l-\omega_k)T-\vartheta_k\big)+\sin\vartheta_k}{\nu_l-\omega_k}
\Bigg] \nonumber\\[0.25em]
&\qquad
+\omega_j\nu_l\sin\vartheta_j\Bigg[
\frac{\cos\vartheta_k-\cos\!\big((\nu_l+\omega_k)T+\vartheta_k\big)}{\nu_l+\omega_k}
+
\frac{\cos\vartheta_k-\cos\!\big((\nu_l-\omega_k)T-\vartheta_k\big)}{\nu_l-\omega_k}
\Bigg]
\Bigg\} \ .
\end{align}
\label{eq:dl_cosines}
\end{widetext}

\section{Optimization with $\pi$-pulses}
\label{sec:pi_opt}

This appendix formulates the optimization problem of combining global
$\pi$-pulses with gradient–modulation segments in order to realize a target
Ising interaction
\begin{align}
\sum_{j,k} \boldsymbol{\varLambda}_{jk} Z_j Z_k \ .
\end{align}

Consider a sequence of $M$ gradient–modulation segments. In segment $p$, with
duration $T_p$ and drive $f_p(t)$, the magnetic-field gradient generates an
effective qubit–qubit matrix $\Lambda^{(p)}$. As discussed in
Eq.~(9) of the main text, in the absence of $\pi$-pulses this matrix can be
written as
\begin{align}
\Lambda^{(p)} = \eta D^{(p)} \eta^\top,
\end{align}
with $D^{(p)}$ diagonal and $\eta$ the normal-mode participation matrix.

Instantaneous $\pi$-pulses on selected ions before and after this
free-evolution block are represented by a diagonal sign matrix
\begin{align}
O^{(p)} = \mathrm{diag}\!\big(s^{(p)}_1,\dots,s^{(p)}_N\big),
\qquad s^{(p)}_j\in\{\pm1\} \ ,
\end{align}
so that the interaction generated in segment $p$ becomes
\begin{align}
\Lambda^{(p)}
= \bigl(O^{(p)}\eta\bigr)\, D^{(p)} \,\bigl(O^{(p)}\eta\bigr)^{\top} \ .
\end{align}
A sequence of $M$ such blocks produces the cumulative interaction
\begin{align}
\Lambda
=
\sum_{p=1}^{M}
\bigl(O^{(p)}\eta\bigr)\, D^{(p)} \,\bigl(O^{(p)}\eta\bigr)^{\top} \ ,
\label{eq:Lambda_eff_appendix_generic}
\end{align}
and the design task is to choose the waveforms $f_p(t)$, the time windows
$T_p$ (which fix $D^{(p)}$), and the sign patterns $O^{(p)}$ so that
$\Lambda$ reproduces the desired target matrix $\boldsymbol{\varLambda}$ (up to irrelevant diagonal terms),
\begin{align}
\Lambda_{jk} = \boldsymbol{\varLambda}_{jk} 
\qquad (j\neq k) \ .
\end{align}

\subsection{Iterative construction}

The single–segment structure above immediately yields a simple diagnostic.
A target interaction $\boldsymbol{\varLambda}$ is exactly realizable with a
\emph{single} global field $f(t)$ and no $\pi$-pulses if and only if
\begin{align}
\boldsymbol{D} = \eta^\top \boldsymbol{\varLambda} \eta
\label{eq:tildeLambda}
\end{align}
is diagonal; in that case one can set $D^{(1)} = \boldsymbol{D}$ and optimize a single drive to implement the corresponding mode-dependent
couplings.

When $\boldsymbol{D}$ has off–diagonal entries, the target lies
outside this single–segment manifold and one must include $\pi$-pulses. A
realization of $\boldsymbol{\varLambda}$ with $M$ segments then amounts to a
decomposition of the form in Eq.~\eqref{eq:Lambda_eff_appendix_generic} with $\Lambda = \boldsymbol{\varLambda}$. 

Rather than assuming \emph{a priori} knowledge of the minimal $M$, it is
convenient to view Eq.~\eqref{eq:Lambda_eff_appendix_generic} as the endpoint of an iterative construction. The procedure starts from the full target
interaction,
\begin{align}
\varepsilon^{(0)} = \boldsymbol{\varLambda} \ ,
\end{align}
and at step $j \ge 1$ selects a sign pattern $O^{(j)}$ and a diagonal matrix
$D^{(j)}$ that extract a diagonal contribution from the current residual
$\varepsilon^{(j-1)}$. In the rotated basis defined by $O^{(j)}\eta$ one writes
\begin{align}
\bigl(O^{(j)}\eta\bigr)^{\top}
\varepsilon^{(j-1)}
\bigl(O^{(j)}\eta\bigr)
= D^{(j)} + \tilde{\varepsilon}^{(j)} \ ,
\end{align}
where $D^{(j)}$ denotes the diagonal part and $\tilde{\varepsilon}^{(j)}$ is
the remaining off–diagonal contribution in that rotated basis. The interaction
generated in segment $j$ is then
\begin{align}
\Lambda^{(j)} = \bigl(O^{(j)}\eta\bigr)\, D^{(j)} \,\bigl(O^{(j)}\eta\bigr)^{\top} \ ,
\end{align}
and the residual in the physical basis is updated according to
\begin{align}
\varepsilon^{(j)} = \varepsilon^{(j-1)} - \Lambda^{(j)} \,.
\end{align}

If at some step $j$ we reach $\varepsilon^{(j)} = 0$, the procedure terminates
with a $j$-segment realisation of $\boldsymbol{\varLambda}$ of the form
Eq.~\eqref{eq:Lambda_eff_appendix_generic}. If $\varepsilon^{(j)}$ remains non-zero, one
proceeds to the next segment, choosing a new sign pattern $O^{(j+1)}$ and
repeating the construction.

\subsection{Numerical strategy}

In practice, the decomposition Eq.~\eqref{eq:Lambda_eff_appendix_generic} is not constructed
symbolically but searched for numerically. The optimisation naturally organises
into two nested loops: a discrete outer search over sign patterns
$\{O^{(p)}\}$ and an inner continuous optimisation over the driving functions
$f_p(t)$ and durations $T_p$.

A convenient procedure is as follows:
\begin{itemize}
    \item \textbf{Single–segment test.}  
    First, test whether the target matrix $\boldsymbol{\varLambda}$ can be realised with a single global field $f(t)$ and no $\pi$-pulses. This amounts to evaluating the diagonalisability condition above: compute Eq.~\eqref{eq:tildeLambda} and check whether its off–diagonal entries are negligible. If so, set $D = \boldsymbol{D}$ and solve an optimisation problem for $\{f(t), T\}$.
    \item \textbf{Multi–segment search.}  
    If the single–segment condition fails, increase the number of segments and perform a joint search over sign patterns and waveforms. Starting from $M=2$, select a family of candidate sign patterns $(O^{(1)},O^{(2)})$ and, for each choice:
    \begin{enumerate}
    \item optimise the first segment by adjusting $\{f_1(t), T_1\}$ so as to minimise a cost functional that measures the off–diagonal residual
    $\varepsilon^{(1)} = \boldsymbol{\varLambda} - \Lambda^{(1)}$;
    \item with $\Lambda_1$ fixed, optimise the second segment $\{f_2(t), T_2\}$ to reduce the remaining residual $\varepsilon^{(2)} = \boldsymbol{\varLambda} - \Lambda^{(1)} - \Lambda^{(2)}$.
    \end{enumerate}
    Among all candidate pairs $(O^{(1)},O^{(2)})$ explored, retain the one that
    yields the smallest off–diagonal residual. If this residual remains above a prescribed tolerance, increase $M$ (e.g.\ to $M=3$) and repeat the procedure, now optimizing over triples $(O^{(1)},O^{(2)},O^{(3)})$ and the corresponding drives. In general, the process is iterated until the convergence criterion $\|\varepsilon^{(M)}\|$ falling below a chosen threshold is satisfied.
\end{itemize}

\section{Optimal $\pi$-pulse design for a uniform Ising coupling under a static gradient}\label{sec:pi-pulses_static}

In this section, the general optimization framework of Appendix~\ref{sec:pi_opt} is specialized to a particularly structured setting: a static magnetic-field gradient and a target uniform all-to-all Ising coupling.

In this scenario, the entangling dynamics in each segment $p$ (Eq.~\eqref{eq:Lambda_eff_appendix_generic}) are fully determined by the normal-mode structure of the chain, so the only continuous control parameters are the segment durations (see Eq.~(22) of the main text). As a result, the cumulative interaction depends \emph{linearly} on these durations and on the sign patterns generated by the $\pi$-pulses. This linear structure allows one to recast the design task as a finite-dimensional convex optimization problem (or, when an exact solution exists, as a linear system), and to determine $\pi$-pulse sequences and segment durations that realize the desired uniform coupling with minimal total gate time.

Under a static gradient, a single free-evolution interval of duration $T$
generates an effective qubit–qubit interaction matrix
\begin{align}
\Lambda_{jk}^{(0)} = -\,T \sum_{l} \nu_l \,\eta_{jl} \eta_{kl} \,,
\end{align}
where $\nu_l$ are the mode frequencies and $\eta_{jl}$ the normal-mode
participation factors. For a sequence of $N_I$ segments of duration $T_p$,
dressed by the $\pi$-pulse patterns $O^{(p)}$, the effective interaction is
\begin{align}
\Lambda
= \sum_{p=1}^{N_I} O^{(p)} \Lambda^{(0)} \bigl(O^{(p)}\bigr)^{\top} \,\tau_p \,,
\label{eq:Lambda_eff_appendix}
\end{align}
with $\tau_p = T_p/T$. Writing
$O^{(p)}=\mathrm{diag}(s^{(p)}_1,\dots,s^{(p)}_N)$ with $s^{(p)}_j\in\{\pm1\}$,
the off–diagonal entries obey
\begin{align}
\Lambda_{jk}
=
\Lambda^{(0)}_{jk}\,\sum_{p=1}^{N_I} s^{(p)}_j s^{(p)}_k\,\tau_p \ .
\end{align}

The goal is to choose sign patterns $O^{(p)}$ and nonnegative durations
$\tau_p$ such that all off–diagonal couplings take a common target value $J$,
\begin{align}
\boldsymbol{\varLambda}_{jk} = J \qquad (j\neq k) \,,
\end{align}
while leaving the diagonal terms unconstrained. This is equivalent to the
family of linear constraints
\begin{align}
\sum_{p=1}^{N_I} s^{(p)}_j s^{(p)}_k\,\tau_p = \lambda_{jk} \qquad (j<k) \,,
\label{eq:pairwise_linear_noQ}
\end{align}
where
\begin{align}
\lambda_{jk} = \frac{J}{\Lambda^{(0)}_{jk}} 
\end{align}
is fixed by the mode structure encoded in $\Lambda^{(0)}$ and by the target coupling strength $J$. 

In principle, solving Eq.~\eqref{eq:pairwise_linear_noQ} requires a
mixed–integer optimisation over the continuous variables $\{\tau_p\}$ and the
discrete choices $\{s_j^{(p)}\}$, as noted in
Appendix~\ref{sec:pi_opt}. In the present setting, this can be avoided by
exploiting two simple observations: (i) the coefficients $\lambda_{jk}$ are
fixed by the underlying problem, and (ii) the constraints depend on each
segment $p$ only through the products $s^{(p)}_j s^{(p)}_k$ and the duration
$\tau_p$.

A $\pi$-pulse configuration $\vec s=(s_1,\ldots,s_N)\in\{\pm1\}^N$ induces
pairwise signs $s_j s_k$ for each $(j,k)$ with $j<k$. Because the global flip
$\vec s\mapsto -\vec s$ leaves these products unchanged, there are only
$M=2^{N-1}$ distinct pairwise sign patterns. Let
$\vec s^{\,(\alpha)}$, $\alpha=1,\dots,M$, be a set of representatives (one
from each pair $\{\vec s,-\vec s\}$), and define the $L\times M$ matrix
\begin{align}
C_{(jk),\alpha} = s^{(\alpha)}_j s^{(\alpha)}_k \ ,
\end{align}
where $L=\binom{N}{2}$ is the number of distinct off–diagonal pairs and the row index $(jk)$ runs over all such pairs with $j<k$.

The next step is to rewrite Eq.~\eqref{eq:pairwise_linear_noQ} in a way that
does not depend explicitly on the number of segments $N_I$, but only on the
distinct sign patterns. Since the constraints involve each segment $p$ only
via $s^{(p)}_j s^{(p)}_k$ and $\tau_p$, any two segments $p$ and $p'$ with the
same configuration $\vec s^{\,(p)} = \vec s^{\,(p')}$ contribute identically to
all pairwise couplings; only their \emph{total} duration matters. It is
therefore natural to group segments by their sign pattern.

For each pattern $\alpha$ the total normalized time spent in that pattern is defined as
\begin{align}
w_\alpha = \sum_{p:\,\vec s^{\,(p)} = \vec s^{\,(\alpha)}} \tau_p  \;\ge 0 \,,
\end{align}
and these are collected into the vector $\vec w=(w_1,\dots,w_M)^{\top}$. Regrouping
the sum in Eq.~\eqref{eq:pairwise_linear_noQ} pattern by pattern then yields
\begin{align}
\sum_{p=1}^{N_I} s^{(p)}_j s^{(p)}_k\,\tau_p
= \sum_{\alpha=1}^M C_{(jk),\alpha}\,w_\alpha \ .
\end{align}
Finally, we collect the $L=\binom{N}{2}$ constraints into a vector form by
defining
\begin{align}
\vec{\lambda}
= \bigl(\lambda_{12},\lambda_{13},\ldots\bigr)^{\top} \ ,
\end{align}
ordered consistently with the rows $(jk)$ of $C$. The uniform–coupling
condition can then be written compactly as
\begin{align}
C\,\vec w = \vec{\lambda} \ , \qquad \vec w \ge 0 \ .
\label{eq:Cw_lambda_uniform}
\end{align}
Thus, for a fixed target strength $J$, an \emph{exact} uniform interaction is
feasible if and only if there exists a nonnegative vector $\vec w$ satisfying
Eq.~\eqref{eq:Cw_lambda_uniform}. In this formulation the explicit number of
segments $N_I$ is irrelevant: only the aggregated durations $w_\alpha$ per
pattern enter the effective coupling, and the search over
$\{\tau_p,\vec s^{\,(p)}\}$ has been reduced to a linear program in $\vec w$.

If Eq.~\eqref{eq:Cw_lambda_uniform} has no solution, the closest achievable
uniform coupling (in the least–squares sense) can be obtained from the
nonnegative least–squares problem
\begin{align}
\min_{\vec w\ge 0}\ \bigl\|C\,\vec w - \vec{\lambda}\bigr\|_2^2 \,.
\label{eq:global_nnls_simplified}
\end{align}
When Eq.~\eqref{eq:Cw_lambda_uniform} \emph{is} feasible, one may further
single out, among the (typically many) exact solutions, the one with minimal
total normalised duration by solving
\begin{align}
\min_{\vec w\ge 0}\ \mathbf{1}^{\top}\vec w
\quad\text{subject to}\quad
C\,\vec w = \vec{\lambda} \,,
\label{eq:min_time_lp_simplified}
\end{align}
where $\mathbf{1}$ is the column vector of all ones, so that the physical
total time is $T_{\mathrm{total}} = T\,\mathbf 1^\top \vec w$. Eq.~\eqref{eq:min_time_lp_simplified} is a linear program, so standard LP solvers either return a globally optimal solution for the minimal total duration or certify that no exact nonnegative $\vec w$ exists.

For the specific case discussed in the main text with $N=4$ ions, there are
$L=6$ off–diagonal pairs in the effective qubit–qubit matrix. Fixing $s_1=+1$
and ordering the pairs as $(12),(13),(14),(23),(24),(34)$ yields
\begin{align}
C \;=\;
\begin{bmatrix}
-1 & -1 & -1 & -1 & +1 & +1 & +1 & +1 \\
-1 & -1 & +1 & +1 & -1 & -1 & +1 & +1 \\
-1 & +1 & -1 & +1 & -1 & +1 & -1 & +1 \\
+1 & +1 & -1 & -1 & -1 & -1 & +1 & +1 \\
+1 & -1 & +1 & -1 & -1 & +1 & -1 & +1 \\
+1 & -1 & -1 & +1 & +1 & -1 & -1 & +1
\end{bmatrix} \, .
\end{align}
Any admissible four-ion sequence is a nonnegative combination of these eight columns. Solving
the linear program in Eq.~\eqref{eq:min_time_lp_simplified} with this $C$ yields the globally shortest exact uniform–coupling schedule for the chosen $J$. In
particular, one finds a minimum total time
\begin{align}
    T = \sum_p T_p \approx 4.30651\,\frac{J}{\eta_C^2}\,\Bigl[\frac{2\pi}{\nu_C}\Bigr] \,.
\end{align}

For reference, in the single–mode limit where only the COM mode is considered, one would require
\begin{align}
  T_{\mathrm{COM}} \;=\; \frac{J}{\eta_C^{2}}\,\left[\frac{2\pi}{\nu_C}\right] \ ,
\end{align}
which reflects that substantial additional evolution is needed to compensate for the inhomogeneous couplings of the rest of the modes.

\section{Gate fidelity}\label{sec:gate_fidelity}

This section derives the expression for the gate fidelity used to benchmark the
protocols via explicit state-vector simulations, for system sizes that remain
numerically tractable.

To quantify the performance of a given protocol, an initial product state
$\rho(0) = \rho_Q \otimes \rho_M$ is considered, where $\rho_Q$ and $\rho_M$
denote the qubit and motional states, respectively. Let $U$ be the ideal (target) unitary gate acting on the qubit degrees of freedom. The ideal final qubit state is then given by
\begin{align}
    \rho_Q(T) = \Tr_M \left[ U \left( \rho_Q \otimes \rho_M \right) U^{\dagger} \right],
\end{align}
where $\Tr_M$ denotes the partial trace over motional modes.

Let $\mathcal{E}(\rho(0))$ represent the actual quantum operation implemented by the protocol. We define the gate fidelity as 
\begin{align}
    F = \frac{1}{d^2} \sum_{\mu, \beta =1}^{d} \sum_{m} \mel{\mu}{ U^{\dagger} \bra{m} \mathcal{E}(\dyad{\mu}{\beta} \otimes \rho_M) \ket{m} U }{\beta} \ ,
    \label{eq:fidelity}
\end{align}
where $d = 2^N$ is the dimension of the $N$-qubit Hilbert space, $\{ \ket{\mu} \}$ and $\{ \ket{\beta} \}$ form an orthonormal basis of qubit states, and $\{ \ket{m} \}$ denotes a truncated basis of the motional Hilbert space. In practice, the motional basis $\{ \ket{m} \}$ must be chosen sufficiently large such that truncation effects remain negligible relative to the dominant error sources in the protocol.

\section{Generation of rainbow states}\label{app:rainbow_generation}

We demonstrate that rainbow states, defined in Eq.~(28) of the main text,
can be generated using only single-qubit gates and the two-qubit Ising
interaction $e^{i g Z_j Z_k}$, which in our setting is implemented
collectively as
\begin{align}
    U(g) = \exp\!\left(i g \sum_{j} Z_j Z_{N-j+1}\right),
\end{align}
starting from the initial product state $\bigotimes_j \ket{0_j}$.
To illustrate the protocol, consider the singlet state
\begin{align}
   \ket{\psi^-}_{j,k}
   = \frac{1}{\sqrt{2}}\left(\ket{0_j}\ket{1_k}
   - \ket{1_j}\ket{0_k}\right)
   \label{eq:singleteq}
\end{align}
between qubits at indices $j$ and $k$.
Products of gates act from right to left throughout this section.

In order to define a sequence of operations that generates
$\ket{\psi^-}_{j,k}$, it is convenient to introduce the Hadamard gate
and the CNOT gate. The Hadamard gate $H_i$ on qubit $i$ is defined as
\begin{align}
    H_{i} = \frac{1}{\sqrt{2}}
    \bigl(\dyad{0_i}{0_i}+\dyad{0_i}{1_i}
    +\dyad{1_i}{0_i}-\dyad{1_i}{1_i}\bigr),
\end{align}
which can also be written in terms of Pauli rotations as
$H_i = e^{i\pi/2} R_{Xi}(\pi)R_{Yi}(\pi/2)$, where
$R_{Xj}(\theta) = e^{-i\theta X_j/2}$ and
$R_{Yj}(\theta) = e^{-i\theta Y_j/2}$ are rotations about the $X$-
and $Y$-axes, respectively.
The two-qubit CNOT gate $\mathrm{CNOT}_{j,k}$ acts on computational
basis states as
$\mathrm{CNOT}_{j,k}\ket{a_j}\ket{b_k} = \ket{a_j}\ket{a \oplus b_k}$,
with $a,b\in\{0,1\}$.

With these definitions, the sequence $Z_j X_k \mathrm{CNOT}_{j,k} H_j$
prepares the singlet state in Eq.~\eqref{eq:singleteq}. Indeed,
\begin{align}
    &Z_j X_k \mathrm{CNOT}_{j,k} H_j \ket{0_j}\ket{0_k} \nonumber \\
    &\quad = Z_j X_k \mathrm{CNOT}_{j,k}
    \frac{1}{\sqrt{2}}\bigl(\ket{0_j}\ket{0_k}
    + \ket{1_j}\ket{0_k}\bigr) \nonumber \\
    &\quad = Z_j X_k \frac{1}{\sqrt{2}}
    \bigl(\ket{0_j}\ket{0_k} + \ket{1_j}\ket{1_k}\bigr) \nonumber \\
    &\quad = Z_j \frac{1}{\sqrt{2}}
    \bigl(\ket{0_j}\ket{1_k} + \ket{1_j}\ket{0_k}\bigr) \nonumber \\
    &\quad = \frac{1}{\sqrt{2}}
    \bigl(\ket{0_j}\ket{1_k} - \ket{1_j}\ket{0_k}\bigr) \nonumber \\
    &\quad = \ket{\psi^-}_{j,k}\, .
    \label{eq:singlet_derivation}
\end{align}

To extend this protocol to the unitary $U(g)$, we must show that
$\mathrm{CNOT}_{j,k}$ can be realized using the $Z_j Z_k$ coupling
present in $U(g)$. A useful identity is
\begin{align}
   e^{i g X_jX_k} = H_k H_j \, e^{i g Z_jZ_k} \, H_j H_k \ ,
\end{align}
which converts $ZZ$-interactions into $XX$-interactions via conjugation
with Hadamard gates.
For $g = \pi/4$, the gate $e^{i (\pi/4) Z_j Z_k}$ is locally equivalent
to a CNOT; one convenient decomposition, up to a global phase, is
\begin{align}
    \mathrm{CNOT}_{j,k}
    = (\mathbb{1}_j \otimes H_k)\,
      e^{i \frac{\pi}{4} Z_j Z_k}\,
      (S_j \otimes S_k)\,
      (\mathbb{1}_j \otimes H_k) \ ,
\end{align}
where $S_\ell = \mathrm{diag}(1,i)$ is the phase gate on qubit $\ell$.
This shows that $\ket{\psi^-}_{j,k}$ can be generated using single-qubit
gates and the two-qubit gate $e^{i g Z_j Z_k}$.

Since $U(g)$ is the product of such $ZZ$-interactions acting on the
symmetrically paired qubits $(j, N-j+1)$, the above procedure can be
applied to each pair. The symmetric pairing ensures that the resulting
state is the rainbow state $\ket{\Psi_{\text{rainbow}}}$ defined in
Eq.~(28) of the main text.

\section{Bound on worst-case gate fidelity}\label{sec:gate_fidelity_bound}

In this appendix we derive the bound on the worst–case gate fidelity stated in
Eq.~(35) of the main text for the family of entangling gates
\begin{align}
U = \exp\!\Bigl(-i\sum_{j\neq k} \Lambda_{jk} Z_j Z_k\Bigr) \,,
\end{align}
where $\Lambda$ is a real symmetric coupling matrix (diagonal entries contribute
only a global phase and can be set to zero without loss of generality). Let
$\boldsymbol{\varLambda}$ denote the target coupling matrix and $\Lambda$ the
one actually realized, and define the coupling error
\begin{align}
\delta\Lambda = \Lambda - \boldsymbol{\varLambda} \, .
\end{align}
Our aim is to bound the \emph{worst–case} state fidelity
\begin{align}
F_{\min}
= \min_{\ket{\Psi}}
\Bigl|\bra{\Psi}U U_T^\dagger\ket{\Psi}\Bigr|^2 \,,
\end{align}
where $U_T$ is the ideal gate associated with $\boldsymbol{\varLambda}$, in
terms of the operator 2–norm $\|\delta\Lambda\|_2$ of the deviation
$\delta\Lambda$.

Define the coherent error Hamiltonian
\begin{align}
\Delta H
= \sum_{j\neq k} [\delta\Lambda]_{jk} Z_j Z_k \,,
\end{align}
so that $U_{\mathrm{r}} U_{\mathrm{T}}^\dagger = e^{-i\Delta H}$.  The operator $\Delta H$ is diagonal in the computational basis
$\{\ket{\vec s}\}$, where $\vec s = (s_1,\dots,s_N)$ with $s_j \in\{\pm1\}$ is
the vector of $Z$–eigenvalues. Acting on such a basis state we have
$\Delta H \ket{\vec s}
= \lambda(\vec s)\,\ket{\vec s}$ with the real eigenvalues 
\begin{align}
\lambda(\vec s)
&= \sum_{j\neq k} s_j [\delta\Lambda]_{jk} s_k \ .
\end{align}

For an arbitrary input state
\begin{align}
\ket{\psi} = \sum_{\vec s} c_{\vec s}\ket{\vec s},
\qquad \sum_{\vec s} |c_{\vec s}|^2 = 1 \,,
\end{align}
we obtain
\begin{align}
\bra{\psi} e^{-i\Delta H} \ket{\psi} = \sum_{\vec s} p_{\vec s} \, e^{-i\lambda(\vec s)} \,,
\end{align}
where $p_{\vec s} = |c_{\vec s}|^2$. Hence
\begin{align}
\nonumber \Bigl|\bra{\psi} e^{-i\Delta H} \ket{\psi}\Bigr|^2
&= \Bigl| \sum_{\vec s} p_{\vec s} e^{-i\lambda(\vec s)} \Bigr|^2 \\
\nonumber &= \Bigl( \sum_{\vec s} p_{\vec s} \cos\lambda(\vec s) \Bigr)^2
  + \Bigl( \sum_{\vec s} p_{\vec s} \sin\lambda(\vec s) \Bigr)^2 \\
&\ge \Bigl( \sum_{\vec s} p_{\vec s} \cos\lambda(\vec s) \Bigr)^2 .
\label{eq:cos-part}
\end{align}

Let
\begin{align}
\lambda_{\max}
= \max_{\vec s} |\lambda(\vec s)|
= \|\Delta H\|_{\mathrm{op}}
\end{align}
be the operator norm of $\Delta H$. We now assume we are in
the small–error regime
\begin{align}
\lambda_{\max} \le \frac{\pi}{2} \,.
\end{align}
Then every eigenphase lies in the interval
$[-\lambda_{\max},\lambda_{\max}]$, on which $\cos\theta$ is non–negative and
monotonically decreasing in $|\theta|$. Hence, for all $\vec s$,
\begin{align}
\cos\lambda(\vec s) \;\ge\; \cos\lambda_{\max} \,.
\end{align}
Using $\sum_{\vec s} p_{\vec s} = 1$, we obtain
\begin{align}
\sum_{\vec s} p_{\vec s} \cos\lambda(\vec s)
\;\ge\; \cos\lambda_{\max}\sum_{\vec s} p_{\vec s}
= \cos\lambda_{\max} \,.
\end{align}
Inserting this into Eq.~\eqref{eq:cos-part} gives the fidelity bound for any
fixed input state,
\begin{align}
\Bigl|\bra{\psi} e^{-i\Delta H} \ket{\psi}\Bigr|^2
\;\ge\; \cos^2\lambda_{\max} \,.
\end{align}
Since this lower bound is independent of the choice of $\ket{\psi}$, it also
bounds the worst–case fidelity 
\begin{align}
F_{\min} \ge \cos^2\lambda_{\max} \,.
\end{align}
In practice, one can either compute
$\lambda_{\max}$ exactly (for small $N$, by evaluating
$\lambda(\vec s)$ over all strings $\vec s$), or upper–bound it in terms of the
2-norm of $\delta\Lambda$. Taking absolute values and using the definition of the operator 2–norm,
\begin{align}
\nonumber  |\lambda(\vec s)|
&= \|\vec s\|_2^2 \,\bigl| \hat s^\top \delta\Lambda \,\hat s \bigr| \\
\nonumber  &\le \|\vec s\|_2^2 \,\max_{\|x\|_2 = 1} |x^\top \delta\Lambda x| \\
\nonumber  &= \|\vec s\|_2^2 \, \|\delta\Lambda\|_2  \\
&= \sqrt{N} \|\delta\Lambda\|_2
\end{align}
where we used $\hat s = \vec s / \|\vec s\|_2$ and $\|\vec s\|_2 = \sqrt{N}$.
Combining this with the bound above yields the practical estimate
\begin{align}
F_{\min}
\;\ge\; \cos^2\!\bigl(N\,\|\delta\Lambda\|_2\bigr) \ ,
\label{eq:fmin_der}
\end{align}
valid whenever $N\,\|\delta\Lambda\|_2 \le \pi/2$.

\begin{figure*}[tp]
\centering
\resizebox{0.99\textwidth}{!}{%
\begin{quantikz}[column sep=0.12cm, row sep=0.14cm, line width=0.4pt]
\lstick{$q_1$}
  & \gate{H}
  & \gate{R_x(\pi)}
  & \gate[wires=4]{U_{\Lambda^{(1,1)}}} 
  & \gate{R_x(-\pi)}
  & \gate[wires=4]{U_{-\Lambda^{(1,1)}}}
  & \gate{R_z(\theta_1 + \theta_2+\theta_3)}
  & \qw
  & \qw
  & \qw
  & \qw
  & \qw
  & \qw
  & \qw
  & \qw
  & \qw
  & \qw
  & \qw
  & \qw
  & \qw
  & \qw
  & \qw
  & \qw
  & \swap{3}
  & \qw
\\
\lstick{$q_2$}
  &\qw
  &\qw
  &\qw
  &\qw
  &\qw
  & \gate{R_z(\theta_1)}
  & \gate{H}
  & \gate{R_x(\pi)}
  & \gate[wires=3]{U_{\Lambda^{(2,1)}}} 
  & \gate{R_x(-\pi)}
  &\qw
  & \gate[wires=3]{U_{-\Lambda^{(2,1)}}}
  & \gate{R_z(\theta_1 + \theta_2)}
  &\qw
  &\qw
  & \qw
  & \qw
  & \qw
  & \qw
  & \qw
  & \qw
  & \swap{1}
  & \qw
  & \qw
\\
\lstick{$q_3$}
  &\qw
  &\qw
  &\qw
  &\qw
  &\qw
  & \gate{R_z(\theta_2)}
  & \qw
  & \qw
  & \qw
  & \qw
  & \qw
  & \qw
  & \gate{R_z(\theta_1)}
  & \gate{H}
  & \gate{R_x(\pi)}
  & \gate[wires=2]{U_{\Lambda^{(3,1)}}} 
  & \gate{R_x(-\pi)}
  & \gate[wires=2]{U_{-\Lambda^{(3,1)}}}
  & \gate{R_z(\theta_1)}
  & \qw
  & \qw
  & \targX{}
  & \qw
  & \qw
\\
\lstick{$q_4$}
  &\qw
  &\qw
  &\qw
  &\qw
  &\qw
  & \gate{R_z(\theta_3)}
  & \qw
  & \qw
  & \qw
  & \qw
  & \qw
  & \qw
  & \gate{R_z(\theta_2)}
  & \qw
  & \qw
  & \qw
  & \qw
  & \qw
  & \gate{R_z(\theta_1)}
  & \gate{H}
  & \qw
  & \qw
  & \targX{}
  & \qw
\end{quantikz}}
\caption{Circuit implementing the four-qubit QFT. Single-qubit gates are Hadamards $H$, $\pi$-pulses $R_x(\pm \pi)$, and $Z$ rotations $R_z(\theta_m)$ with $\theta_m=\pi/2^{m+2}$ for $m=1,2,3$. The multi-qubit blocks $U_{\pm\Lambda^{(\ell,p)}}$ implement, at each stage $\ell$, the effective $ZZ$–coupling matrix $\boldsymbol{\varLambda}^{(\ell)}$ defined in Eq.~(33) of the main text. In our construction, the target interaction for stage $\ell$ is realized by combining the blocks $U_{\Lambda^{(\ell,p)}}$ and $U_{-\Lambda^{(\ell,p)}}$ with appropriate $\pi$-pulse patterns, so that $\boldsymbol{\varLambda}^{(\ell)} \simeq O^{(1)}\Lambda^{(\ell, 1)}\left(O^{(1)}\right)^{\top} - \Lambda^{(\ell, 1)}$ with $O^{(p)}$ the diagonal sign matrices set by the corresponding $\pi$-pulse patterns.
The final SWAP gates (optional) restore the standard qubit ordering; if the circuit is measured immediately these SWAPs may be omitted and the resulting bit reversal handled in classical post-processing, otherwise they can themselves be implemented using the protocol of Sec.~III B of the main text.}
\label{fig:qft_circuit}
\end{figure*}
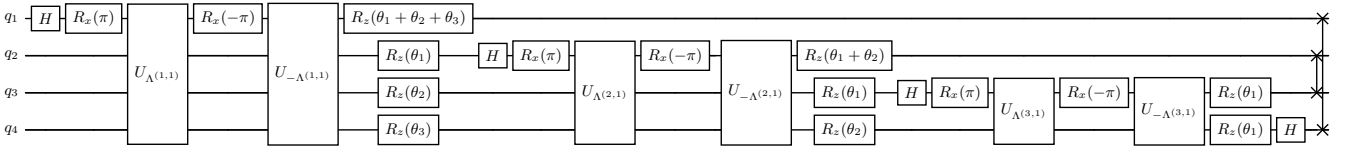

\end{document}